\documentclass{article}

\usepackage{amsmath}
\usepackage{amsfonts}
\usepackage{graphics}

\usepackage[dvips]{epsfig}
\usepackage{float}
\usepackage{latexsym}

\def\vonkarman{von K\'{a}rm\'{a}n\ }
\def\crossstar{$\ast$}
\def\calR{\mathcal{R}}
\def\calL{\mathcal{L}}
\def\mybox{{\tiny $\Box$}}
\def\universal{20 \%}
\def\oneplusuniversal{1.2}

\newcommand{\legend}[1]{\parbox{6in}{\small\vspace{0.3\baselineskip} #1}}

\begin{document}

\title{Scaling of the buckling transition of ridges in thin sheets}
\author{Brian DiDonna\\
Department of Physics, University of Chicago, Chicago, IL 
60637}
\date{\today}



\maketitle

\begin{abstract}
When a thin elastic sheet crumples, the elastic energy condenses into a
network of folding lines and point vertices.  These folds and vertices
have elastic energy densities much greater than the surrounding areas,
and most of the work required to crumple the sheet is consumed in
breaking the folding lines or ``ridges''. To understand crumpling it is
then necessary to understand the strength of ridges.  In this work, we
consider the buckling of a single ridge under the action of inward
forcing applied at its ends. We demonstrate a simple scaling relation
for the response of the ridge to the force prior to buckling. We also
show that the buckling instability depends only on the ratio of strain
along the ridge to curvature across it. Numerically, we find for a wide
range of boundary conditions that ridges buckle when our forcing has
increased their elastic energy by $\universal$ over their resting state
value.  We also observe a correlation between neighbor interactions and
the location of initial buckling. Analytic arguments and numerical
simulations are employed to prove these results.  Implications for the
strength of ridges as structural elements are discussed.
\end{abstract}

%
%
\section{Introduction}

%
%

The crumpling of a thin sheet is a phenomenon which we encounter every
day, yet the equations governing crumpled systems are almost
completely intractable without the introduction of drastic simplifying
assumptions\footnote{As Landau and Lifshitz dryly put it,
(translated)~\cite{landau} ``These equations are very complicated, and
cannot be solved exactly, even in very simple cases.''}~\cite{landau}. 
At the same time, this mundane occurrence exhibits some of the more
intriguing behaviors of modern soft matter physics, such as phase
transitions~\cite{nelson}, scaling~\cite{science.paper}, and 
energy condensation~\cite{eric}.

\begin{figure}

\center

\epsfig{file=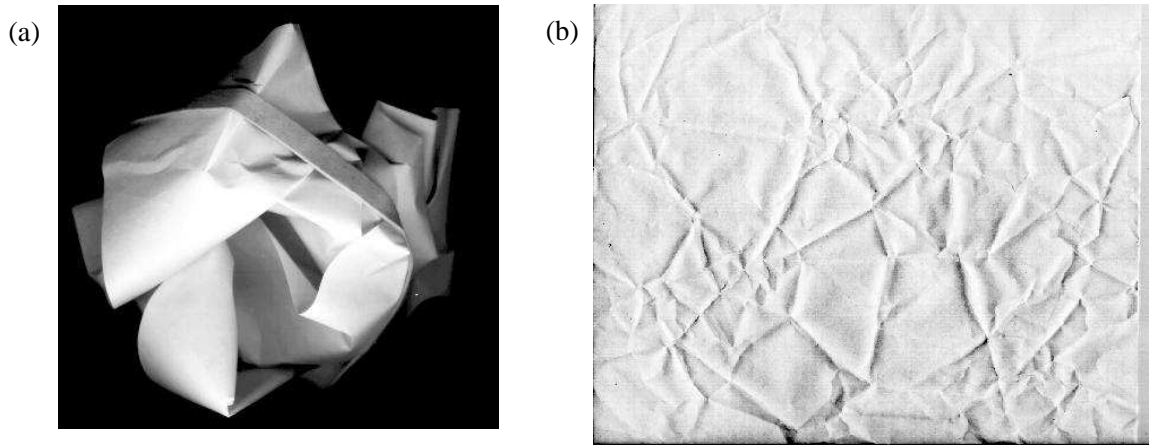}

\caption{A typical crumpled sheet}
\label{fig:paper}

\legend{Image (a) shows a sheet of paper which has been lightly crumpled 
between the hands. Image (b) is the same sheet unfolded -- lines and points resulting from plastic deformation show the former
locations of folds and vertices in the crumpled state. Image courtesy
the authors of~\cite{science.paper}.}

\end{figure}

For a large class of compressive
boundary conditions, the energetically preferred configurations
of crumpled thin sheets
consist of mostly flat regions bounded by straight folds and
point-like vertices. Figure~\ref{fig:paper} shows an example of the resulting network of folds and points in a crumpled sheet of paper.
One approach to analyzing such configurations is to
treat them as patches of unstrained
surface bounded by regions
of discontinuous curvature~\cite{KJ}. 
Boundary layer solutions are then grafted to
the regions of sharp curvature, and the total energies of the
configurations are compared to find local or global energy minima.

In the last several years the structure and energy of  
boundary layer solutions around straight folds and isolated vertices
in crumpled sheets
have been studied in 
detail, both from a physical 
perspective~\cite{nelson,our.stuff,eric,Alex,
science.paper,Pomeau,Pomeau2,maha,boudaoud,chaieb.cone,
maha.cone,chaieb.crescent,chieb.creases,
chaieb.creases.exp,Cerda,anomylous,tanizawa,ajay,GAOC}, and
mathematical perspective~\cite{KJ,sternberg,phase_transitions,
AG,DKMO,ADM,BCDM}. 
Related geometries such as thin film
blistering~\cite{GO,keer,Audoly,JS1,JS2}, 
thin viscous sheets~\cite{boudaoud.viscous},
thin film actuators~\cite{james},
molecular sheets~\cite{Haoli.TW,fullerenes},
and the generalization of crumpling to higher 
dimensions~\cite{eric.math,eric,shankar.witten,didonna.shankar}
have also received attention. 
In particular, the boundary layer around
a fold was extensively studied by 
Lobkovsky and Witten~\cite{our.stuff,Alex,science.paper}. They
called the energetically preferred configuration a ``stretching ridge,''
since it comes about through the balance of bending and stretching energy
on the fold line, where both energies are of comparable magnitude.

By viewing a crumpled sheet as a collection
of ridges and vertices and adding up the known energetic cost of each
unit we may arrive at a reasonable estimate of the total elastic energy
in a crumpled sheet. However, part of the picture is still missing.
We know from common experience that the crumpling of a piece of paper
between the hands is a dynamic process, with the details of the final
shape of the paper depending strongly on the history of applied forces
and the effects of geometric frustration. In order to understand highly
crumpled objects, which are clearly not free to find a global minimum of
elastic energy, we need to know more about the energetic paths the
membrane may take from one crumpled state to another. 

This thesis investigates the energetic pathway whereby one
ridge buckles into several under the action of a compressive
load. The work builds on Lobkovsky's scaling analysis of stretching
ridges, though we differ in our treatment of applied forces at the tips
of the ridge. Using improved simulational techniques and greater
computing power, we investigate the buckling
transition in far more detail than was previously possible.
We then analyze the transition in a framework of stability and
bifurcation theory, comparing and contrasting the transition on the
boundary layer to the well studied subject of thin cylinder stability.

We begin in Section~\ref{ch:prebuckle}
by reviewing the elastic theory of thin sheets,
giving a brief derivation of the \vonkarman equations upon which our
analysis is based. We then present Lobkovsky's derivation of ridge
scaling, and his treatment of small perturbations to resting ridges. 
He chose a perturbative approach which assumed linear response to 
applied forces. After describing his method, we present an 
alternate approach which integrates applied forces into 
the original scaling equations. Our new approach provides better 
descriptive power when considering some special cases of applied 
forcing, since it does not assume that the applied forces are small. 
We then argue that our technique applies well to the highly relevant 
case of a ridge with inward point forces applied at its tips. Force 
scaling exponents are derived for this case.

\begin{figure}

\center

\epsfig{file=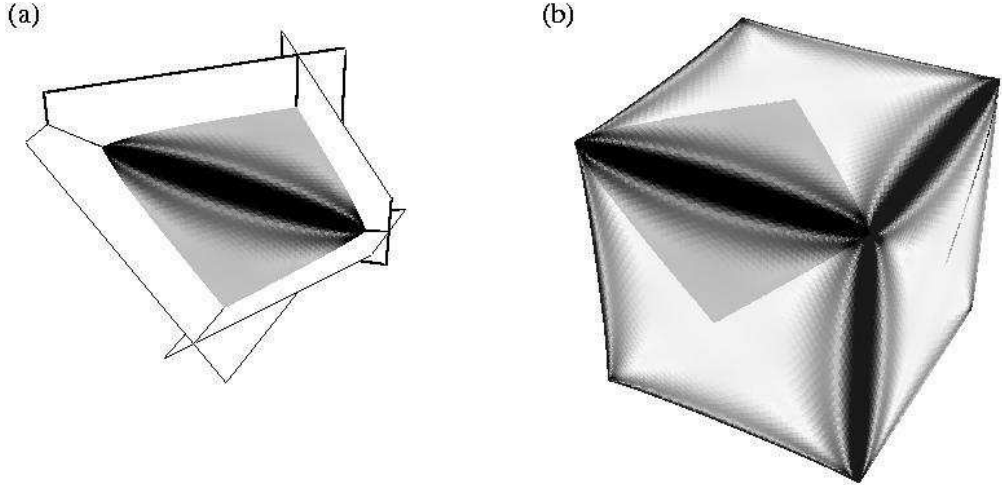, width=5.25in}

\caption{Typical elastic sheet used in this study}
\label{fig:grid}

\legend{(a) The resting configuration of the simulated sheet 
with no external forces. It also shows the reflection planes to 
which the sheet edges are constrained. (b) How the 
simulated sheet is equivalent to one edge of a cube, when the mirror 
images of the sheet across the reflective planes are drawn in.
The thickness of the sheet is $.0004$ of 
the the edge length and 
the Poisson ratio is $1/3$.  Darker shading represents 
higher strain energy density. The entire simulated sheet is
uniformly darkened to distinguish it from its mirror images in (b).  
Slight numerical symmetry breaking between 
the left and right sides  of the diamond created slight 
mismatches of the inferred surfaces on other faces, such as the right-hand 
face.  The numerical grid is visible as a quilt-like texture.  It has a finer 
scale at the edges and corners where curvature is larger.}

\end{figure}

In Section~\ref{ch:sims}
we present numerical evidence to support our 
scaling arguments. We devised a novel finite 
element program to increase the accuracy and efficiency 
with which we can simulate elastic sheets.
This data was also presented in a companion 
paper~\cite{anomylous}.
For most simulations presented here, the shape and 
boundary conditions of the grid were chosen to simulate a section 
of a cubical box, as shown in Figure~\ref{fig:grid}. The grid covers 
one edge of the cube and has reflective boundary conditions.
Simulating only this edge is equivalent to simulating a cube which is 
constrained to have $16$-fold symmetry (all fold-lines equivalent).
We argue that the reflective boundary conditions are representative 
of the real boundary conditions for a single ridge in a general 
crumpled sheet, since the forces which maintain the angle of a given 
ridge are most often exerted by the surrounding ridges. In later 
simulations we change the resting angle of the ridges, so that they do 
not correspond to edges of any closed polyhedral surface -- the initial 
choice of the $\frac{\pi}{2}$ dihedral
angle corresponding to a cubic surface was 
arbitrary. Using data generated by these simulations, we demonstrate 
a scaling solution for ridges with inward
forces applied at their ends. We also provide numerical evidence that 
the critical strain and curvature on the ridge at the buckling threshold 
scale with the same exponents as for ridges at rest.

Finally, in Section~\ref{ch:buckling}
we consider the buckling transition. When the ridge is 
subject to strong enough forcing at its endpoints, it breaks into 
several ridges as shown in Figure~\ref{fig:buckled}. 
We show that this 
transition is identical to the 
bifurcation by which thin cylinders buckle. We begin by reviewing the 
buckling transition of a cylinder under uniform axial compression. 
Thin cylinders subject to such forcing are observed to buckle in a regular
diamond shaped pattern with azimuthal periodicity determined by their 
thickness and radius. The first bifurcation is shown to occur at a 
critical stress which is inversely proportional to the radius of the 
cylinder. We apply the relations derived for the cylinder to the 
geometry of the ridge boundary layer and show that it is consistent 
with the observed scaling of the ridge buckling transition. We also 
show that the small longitudinal curvature along the 
ridge and the non-uniformity of the curvature and strain on 
the ridge have only a weak effect on the buckling transition derived 
for the uniform cylinder.

Adopting the hypothesis that the cylinder buckling mode accurately 
describes the buckling of stretching ridges, we make two previously 
untested predictions for the buckling behavior of ridges.
Our first prediction concerns the normal mode 
which is associated with the buckling motion.
We reanalyze our existing data from Section~\ref{ch:sims} to isolate
the soft normal mode which becomes unstable as its 
associated spring constant passes through zero. Comparing the
numerical results to the theory developed for the cylinder, we show that 
the rate at which the spring constant approaches zero with increasing 
stress along the ridge is on the same order of magnitude as for the 
normal mode associated with cylinder buckling. Our other 
prediction is that, for a given thickness to ridge length ratio, the 
buckling transition will occur at the same maximum ratio of stress to 
curvature. To show this, we simulate several variations on the ridge 
geometry described above. Although the different geometries buckle at 
different values of longitudinal stress and transverse curvature, they
all buckle near the same value of the ratio of these parameters.

We conclude in Section~\ref{ch:discussion} by discussing the
implications of this research for the strength of ridges as load bearing
objects. The immediate consequence of our research is that ridges are
not as strong as once thought. Since the ridge buckles when the total
longitudinal stress along the ridge-line reaches a critical value, the
pre-existing stress found in resting ridges makes them easier to break
than stress free shells and cylinders with the same radius. 
Strategies for strengthening
ridges are discussed, along with topics for future research concerning
the buckling transition.

\section{Pre-buckling behavior of ridges}
\label{ch:prebuckle}

\subsection{The \vonkarman equations}
\label{sec:vonkarman}

We consider an 
idealized thin elastic sheet with uniform
Young modulus $Y$. The sheet has constant
thickness $h$ which is much smaller than its spatial extent in the
other two material directions. In the regime where
elastic distortions
are small and slowly varying on the scale of $h$,
stresses along the thin
direction can be neglected in comparison to those in the long
directions.
In this regime the thin direction can then 
be integrated out of the governing
elastic equations~\cite{landau}, so that
the sheet is completely characterized
by its $2$-dimensional center surface. 

We assume that our sheet has no intrinsic 
strain or curvature, so we may define on it
material coordinates $\vec{x} \subset \calR^2$. 
The embedding of the sheet into 
$\calR^3$ can then be expressed as some function
$\vec{r}(\vec{x})$ of the material coordinates $\vec{x}$. 
The strain is defined as the change in length
element $dl$ under the embedding:
\begin{equation}
{dl'}^2 = dl^2 + 2 \gamma_{ij} dx_i dx_j,
\end{equation}
so that 
\begin{equation}
\gamma_{ij} = \frac{\partial \vec{r}}{\partial x_i} \cdot
\frac{\partial \vec{r}}{\partial x_j} - \delta_{ij}.
\label{eq:straindef}
\end{equation}


The most general quadratic form for the stretching elastic energy 
density in terms of the strain can be written as
\begin{equation}
\calL_S= \frac{Yh}{2(1-\nu^2)}
\left( \gamma_{ij}\gamma_{ij} + \nu
\epsilon_{ik} \epsilon_{jl} \gamma_{ij} \gamma_{kl}\right),
\label{eq:stretchen}
\end{equation}
where $\nu$ is the Poisson ratio and $\epsilon_{ij}$ is the
antisymmetric tensor.

The stress is defined as the variation of $\calL_S$ with strain,
$\sigma_{ij} = \partial \calL_S/\partial \gamma_{ij}$,
so we can write $\calL_S = \frac{1}{2} 
\sigma_{ij} \gamma_{ij}$ with
\begin{equation}
\sigma_{ij} = \frac{Yh}{1-\nu^2} \left[\gamma_{ij} + \nu \
\varepsilon_{ik} \varepsilon_{jl} \gamma_{kl} \right].
\label{eq:stressstrain}
\end{equation}

Because of the nonzero thickness of the sheet, there is also an energy
cost associated with bending of the sheet out of its local material
plane. Bending is quantified by the extrinsic
curvature, which
can be expressed as the component of the second derivative of
$\vec{r}$ normal to the local material frame.
\begin{equation}
C_{ij} = \frac{\partial \vec{r}}{\partial x_i \partial x_j} \cdot \vec{n}.
\label{eq:curvdef}
\end{equation}
The inverses of the eigenvalues of the curvature tensor are the local
principal radii of curvature of the sheet. Geometric
constraints~\cite{diff_geom,sectional_curvature} 
require that the curvature tensor
satisfy
\begin{equation}
\partial_i C_{jk} = \partial_k C_{ji}
\label{eq:codazzi}
\end{equation}
for a sheet with flat metric.

The most general quadratic form for the
energy density associated with bending can be written
\begin{equation}
\calL_B = \frac{1}{2} \left( \kappa C_{ij}\kappa C_{ij} + 
\kappa_G 
\varepsilon_{ik} \varepsilon_{jl} C_{ij} C_{kl} \right),
\label{eq:benden}
\end{equation}
where the coefficient of $\kappa_G$ is the familiar Gaussian curvature.

To establish the connection between the curvature energy and the bulk
elastic moduli, we introduce the so-called Monge coordinates.
These coordinates locally parameterize the 
center surface of the sheet by
\begin{equation}
\vec{r}(\vec{x}) = ( \ x+u \ , \ y+v \ , \ w \ ),
\end{equation}
where $x$ and $y$ are the material coordinates and $u$, $v$ and $w$ 
are small deviations from the flat, unstrained state.

Microscopic considerations
of the finite sheet thickness~\cite{landau} yield an
energy functional for $w$:
\begin{equation}
\calL_{w} = \frac{Yh^3}{24(1-\nu^2)}
\left[ \left( \partial_i \partial_j w \right)^2 + 
\nu \varepsilon_{ik} \varepsilon_{jl} 
\left( \partial_i \partial_j w \right)
\left( \partial_k \partial_l w \right) \right].
\end{equation}
To lowest order in $w$, Eq.(~\ref{eq:curvdef}) gives
$C_{ij} = \partial_i \partial_j w$, so we can immediately make the
identification $\calL_{B} = \calL_{w}$ with
\begin{gather}
\kappa = \frac{Yh^3}{12(1-\nu^2)}, \nonumber \\
\kappa_G = \frac{Yh^3 \nu}{12(1-\nu^2)}.
\label{eq:kappaY}
\end{gather}

The equations of equilibrium can be found by setting the variation of
the total energy of the sheet to zero. In the presence of an external
pressure field $P$, the condition for an energy extremum becomes
\begin{equation}
\delta \left[ \int \calL_S  + \int \calL_{w} \right] +
\int P \delta w = 0
\label{eq:variation}
\end{equation}
Grouping terms for in and out-of-plane displacements 
yields the equilibrium conditions:
\begin{gather}
\partial_j \sigma_{ij} = 0.
\label{eq:eqstrain}
\end{gather}
\begin{gather}
\frac{Yh^3}{12(1-\nu^2)} \nabla^2 \nabla^2 w -
\partial_j \left(\sigma_{ij} \partial_i w \right) = P,
\end{gather}
Therefore the equilibrium condition can be written:
\begin{equation}
\kappa \nabla^2 C_{ii} = \sigma_{jk} C_{jk} + P.
\label{eq:forcevk}
\end{equation}

Equations (\ref{eq:eqstrain}) and (\ref{eq:forcevk}) are not enough to
completely specify the system, so another equation of state is
necessary. An appropriate equation is Gauss' fundamental
theorem of surfaces~\cite{diff_geom,sectional_curvature},
\begin{equation}
\det C_{ij} = \partial_i \partial_j \gamma_{ij} - \nabla^2 \mathrm{tr}
\gamma_{ij},
\label{eq:gauss}
\end{equation}
which relates strain to Gaussian curvature. Together the force
Eqs.~\ref{eq:eqstrain}~and~\ref{eq:forcevk}
along with the constraint Eqs.~\ref{eq:codazzi}~and~\ref{eq:gauss} 
are enough to completely
determine the equilibrium configurations of a sheet up to arbitrary
translations and rotations. 
Equations~\ref{eq:forcevk}~and~\ref{eq:gauss} 
are respectively called
the force and geometric \vonkarman equations~\cite{vK}. 

As a consequence of Eq.~\ref{eq:codazzi}, the curvature tensor
can be written as
the derivative of a continuous
curvature potential
\begin{equation}
C_{ij} = 
\partial_i \partial_j f.
\label{eq:fdef}
\end{equation}
Here the potential $f(\vec{x})$ is not identical to the
local function $w$
used above, but is
approximately equal to it for nearly flat surfaces.
Also, Eq.~\ref{eq:eqstrain}
is automatically satisfied if we write $\sigma_{ij}$
in terms of a stress potential $\chi$,
\begin{equation}
\sigma_{ij} = \varepsilon_{ik} \varepsilon_{jl}
\partial_k\partial_l \chi.
\label{eq:chidef}
\end{equation}

In terms of the potentials $\chi$ and $f$, the \vonkarman equations
assume the very compact form
\begin{gather}
\kappa \nabla^4 f = [\chi, f] + P, \\
\frac{1}{Yh} \nabla^4 \chi = -\frac{1}{2} [f, f],
\label{vk2}
\end{gather}
where the brackets product represents
\begin{equation}
\left[ a,b \right] = \epsilon_{\alpha\mu} \epsilon_{\beta\nu}
\left( \partial_\alpha \partial_\beta a \right)
\left( \partial_\mu \partial_\nu b \right),
\end{equation}


\subsection{Ridge scaling solution}
\label{sec:ridgescale}

\begin{figure}

\center

\epsfig{file=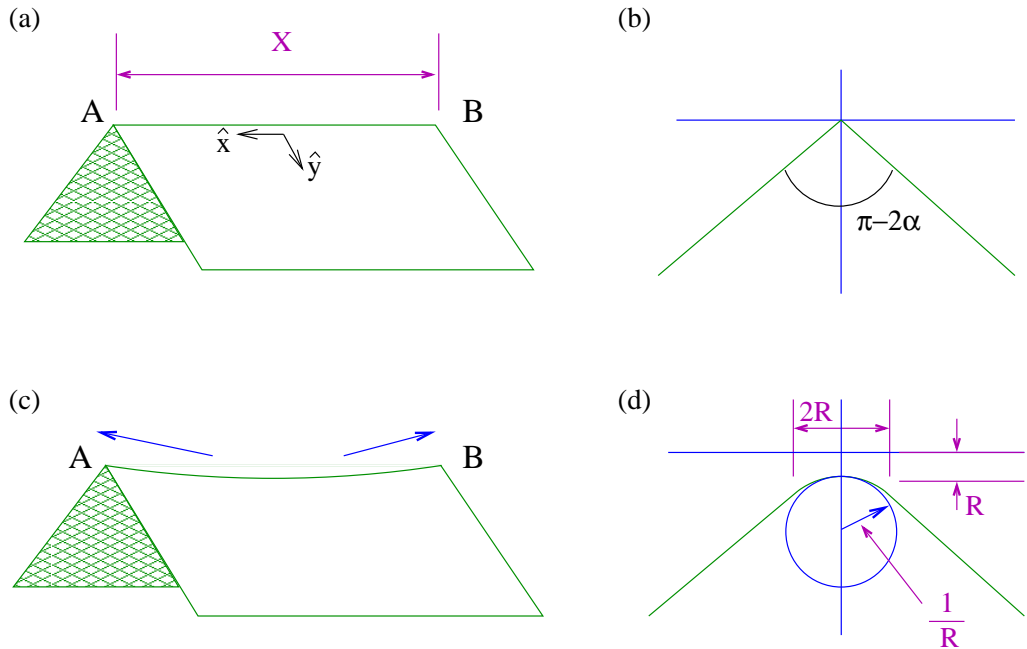}

\caption{The stretching ridge boundary layer}
\label{fig:ridgeexplain}

\legend{Images (a) and (b) show the geometry of a sheet with one sharp fold. 
(a) is a side view, while (b) is a cross section of the fold at the 
midpoint of the sheet. Images (c) and (d) show a representation of 
the ``stretching ridge'' configuration, in which the boundaries of 
the sheet are still required to make a sharp angle. 
Image (d) shows the same cross 
section which is shown in (b) -- this image illustrates how the 
curvature across the fold-line is lessened in the ridge
configuration, and 
the region of large curvature has a width on the order of the maximum 
radius of curvature. Image (c) shows how the softening of the 
curvature requires stretching of the sheet along its mid-line. The 
geometry shown here, used for the arguments presented 
in~\cite{science.paper}, requires extensional strain along the 
mid-line -- other geometries can have a net compressive strain along 
the mid-line, but the driving balance between curvature and strain 
remains the same~\cite{Alex}.}

\end{figure}

We wish to use the \vonkarman equations to study the
the boundary layer around a folding line in an elastic sheet. This 
fold plus boundary layer configuration is what Lobkovsky et al.
termed the
``stretching ridge.''~\cite{science.paper,Alex}
An intuitive picture of the structure of the
boundary layer is presented in Figure~\ref{fig:ridgeexplain}. As we showed
in the last section, for thin sheets the bending elastic modulus is
less than the stretching modulus by a factor of $h^2$. 
Therefore, for very
thin sheets with relatively free boundary conditions\footnote{For a
discussion of the role of boundary conditions, 
see~\cite{didonna.shankar}.}  minimal
energy configurations are mostly strain free, with large deformations
concentrated around folding lines. Very near a folding line, the
curvature approaches a scale where bending and stretching energies are
once again comparable and the local configuration is determined by a
balance between these two energies. 

Conceptually, as  
Figure~\ref{fig:ridgeexplain} presents, the stretching ridge can be
approached from the limit of zero thickness, where the curvature at the
folding line becomes singular. The boundary conditions which
create the fold are point-like vertices at points A and B, which
are maintained at a sharp curvature. 
As the thickness of the sheet increases, it is
energetically favorable for the middle section of the ridge to have a
lower curvature at the expense of stretching along the length of the
ridge. The boundary layer around the fold thus acquires a saddle-like
shape as shown in Figure~\ref{fig:ridgeexplain}(c) and~(d). The width
of the boundary layer is on the same order as the transverse radius of
curvature $R$ and is much less than the length of the ridge $X$.

Now we proceed to apply the \vonkarman equations to the stretching ridge.
This system has two well
defined typical length scales -- the sheet thickness $h$ and the ridge
length $X$. We can therefore rescale the \vonkarman equations into a
more convenient dimensionless form by expressing all lengths in units of
$X$ and all energies in units of $\kappa$. The \vonkarman equations then
become
\begin{gather}
\nabla^4 \bar{f} = [\bar{\chi}, \bar{f}] + \bar{P}, \nonumber \\
\lambda^{2}  \nabla^4 \bar{\chi} = -\frac{1}{2} [\bar{f}, \bar{f}].
\label{vk2s}
\end{gather}
Here $\bar{\chi}$, $\bar{f}$, and $\bar{P}$
represent the stress potential, curvature potential and 
external normal forces in respective natural 
units of $\kappa X^{-2}$, $X$, and $\kappa X^{-3}$.
All derivatives are taken with respect to the dimensionless variables
$x/X$ and $y/X$. The dimensionless small parameter $\lambda$ is 
given by
\begin{equation}
\lambda = \frac{\sqrt{\kappa/Yh}}{X} = 
\frac{1}{\sqrt{12 \left( 1 - \nu^2 \right)}} \left(
\frac{h}{X} \right),
\end{equation}

We consider a sheet with edge boundary conditions sufficient to 
create a single ridge and no
normal forces ($P=0$). We define our coordinate frame so that
the origin is at the center point of the ridge and
the center-line of the ridge is
parallel to the $\hat{x}$ material direction.
Since $\lambda$ comes into the von K\'{a}rm\'{a}n equations
multiplying the stress source term,
the possible configurations of a thin elastic sheet are well
described by a stress free, $\lambda = 0$, folding solution 
plus boundary layers at the fold lines. 
Lobkovsky's insight in~\cite{Alex}
was to try a scaling solution for the boundary layer
of a single ridge 
which matched the $\bar{f}$ scaling of the 
outer, sharp fold solution. For a
fold of dihedral angle $\pi-2 \alpha$ across the line $y=0$, 
$\bar{f} = \alpha \left| y/X \right|$. Accordingly, on the boundary layer
$\bar{f}$ should scale with the same power of $\lambda$
as the dimensionless transverse coordinate $y/X$. He therefore tried a
scaling solution  of the form:
\begin{equation}
\tilde{f} = \lambda^{\beta} \bar{f}, \
\tilde{\chi} = \lambda^{\delta} \bar{\chi}, \
\tilde{y} = \lambda^{\beta} y/X, \
\tilde{x} = x/X,
\label{eq:scalings}
\end{equation}
where the tildes denote dimensionless, scale free coordinates and
functions. Taking $\beta < 0$ gives the proper 
limiting case of sharp curvature at zero thickness. 
Substitution into the rescaled \vonkarman equations 
with $\bar{P}=0$ and 
retention of only the largest terms yields
\begin{gather}
\lambda^{3 \beta} \frac{\partial^4 \tilde{f}}
{\partial \tilde{y}^4} = \lambda^{\beta-\delta} 
[\tilde{\chi}, \tilde{f}], \nonumber \\
\lambda^{2-\delta+4\beta}  
\frac{\partial^4 \tilde{\chi}}
{\partial \tilde{y}^4}
= -\frac{1}{2} \lambda^{0} 
[\tilde{f}, \tilde{f}].
\label{vk2sc}
\end{gather}
In the $\lambda \rightarrow 0$ limit the two sides of each equation must
scale identically. Solving for the exponents yields
\begin{equation}
\beta = - \frac{1}{3}, \ \delta = \frac{2}{3}.
\end{equation}
This translates to 
$\lambda^{1/3}$ scaling of the boundary
layer width, $\lambda^{-1/3}$ scaling of the transverse ridge curvature, 
and $\lambda^{2/3}$ scaling of the strain along the ridge length. 
So, to within factors of order unity, the
radius of curvature across the ridge is $R \approx 
X \lambda^{1/3}$, and the total width of the boundary layer should be
about the same. Also, if we assume most of the elastic energy 
is concentrated on the boundary layer, a region of area 
$\lambda^{1/3} X^2$, then
integration of the energy functionals in Eq.~\ref{eq:stretchen}
and Eq.~\ref{eq:benden} yields total bending and stretching energies
which scale as $\kappa \lambda^{-1/3}$.

\subsection{Response to external forces}

In deriving the ridge scaling exponents,
Lobkovsky et al. 
posited a set of ``minimal'' boundary conditions to create a
ridge -- purely normal forces are applied as necessary at the edges of
the sheet to maintain straight, sharp folds.
Relying on these boundary conditions as necessary, he derived
functional forms for many of the geometric and energetic quantities of
interest on the ridge~\cite{Alex,our.stuff}.
However, the derivation of the generic scaling
exponents presented in
Section~\ref{sec:ridgescale} does not rely on any specific
boundary conditions, and so these 
exponents should apply to a much broader
class of ridge configurations than just the ``minimal'' ridge.

These scaling arguments still place 
a few limitations on generic ridges.
An important ingredient in the ridge scaling derivation is that there
are only two length scales, the thickness $h$ and the ridge length $X$. 
This is the essence of our notion of an unperturbed, resting
ridge -- the boundary conditions do not impose another length scale
on the problem. 

To quantify the response of ridges to external forces, we must 
study how the unperturbed ridge
evolves as a new length scale is added to the system. Our common
experience gazing at a crumpled sheet
of paper tells us that real elastic ridges do not live in isolation, but
are influenced by other ridges around them as well as by the global
geometry of the sheet. Since ridge scaling is witnessed under such
circumstances, the governing equations must be fairly insensitive to
most small perturbations. 

Previous efforts~\cite{our.stuff}
to study the response of ridges to external forcing
relied on a perturbative scheme which assumed linear response to small
forces. The response of the ridge to in-plane 
external forcing at the edges of the sheet was deduced by first solving
for the stress field such a force would produce in a flat sheet, then 
using this solution to modify the equations governing the ridge. 
A ridge scaling
solution was then substituted into the 
modified \vonkarman equations and series
expanded in the magnitude of the external force. This technique had the
weakness that the resulting expressions for the force response had an
undetermined exponent (though the possible values of the exponent were
limited). The technique was also very complicated.

We use a less specific, more intuitive
approach to study the overall response of a
ridge to external forcing. Instead of solving for the detailed
behavior of the ridge strain and curvature as the force is applied,
we focus on how the forcing must be rescaled to have an equivalent
effect on ridges of different length ratios $\lambda$. 
For example,
forces applied normal to the sheet enter the \vonkarman equations
through the term $P$ in the force equation. If we add the 
dimensionless form of this term back
to the ridge scaling form of the equation (Eq.~\ref{vk2sc}), we find
\begin{equation}
\lambda^{-1} \frac{\partial^4 \tilde{f}}
{\partial \tilde{y}^4} = \lambda^{-1} 
[\tilde{\chi}, \tilde{f}] + \bar{P},
\end{equation}
where we have used the ridge scaling 
values for $\beta$ and $\delta$. In order for the solutions
to $\tilde{f}$ and $\tilde{\chi}$ to remain scale invariant, the field
$P$ must obey the scaling form
\begin{equation}
\bar{P} = \lambda^{-1} \tilde{P} 
\left( \tilde{x}, \tilde{y} \right),
\label{eq:ptilde}
\end{equation}
where $\tilde{P}$ is a dimensionless. scale free function of 
$\tilde{x}$ and  $\tilde{y}$. Conversely, the existence of the
scale-free function $\tilde{P}$ gives a relation 
between two equivalent
external forces on ridges of different $\lambda$:
\begin{equation}
\bar{P}_1 = \left( \frac{\lambda_1}{\lambda_2} 
\right)^{-1} \bar{P}_2.
\end{equation}

Thus we assume
that with proper rescaling of our applied forces, the fields $f$ and
$\chi$ obey the same $\lambda$ scaling 
laws  on  forced ridges as on  resting ridge.
This notion allows us to make strong statements about the complex
evolution of ridge shape with applied forcing. Still,
the flexibility of our approach is greatly limited by 
the requirement implicit in Eq.~\ref{eq:ptilde}
that the spatial dependence of the applied force
also scales\footnote{This limitation is
similar to Lobkovsky et al.'s requirement that his forcing be ``compatible''
with the boundary conditions for his ``minimal'' ridge. However, he
still had complete freedom to study any form of boundary 
forcing in the material
plane, since the minimal ridge boundary conditions only specify the
normal forces.}  with $\lambda$.
A perturbation scheme with which  we are not free to fix the
location of our perturbing force seems to be of limited physical
interest. However, this scheme is sufficient to study some special cases
which are particularly important. For example, it is well suited to the
problem of a ridge with external point forces applied at its vertices,
since the spatial location of the equivalent 
forcing will clearly remain fixed as
$\lambda$ is varied. The other benefit of this scheme is that it does
not rely on assumptions of linear response, so it
describes the force response over changes of order unity in the fields
$f$ and $\chi$.

\subsubsection{Point forces applied at ridge vertices}
\label{sec:tipforcing}

External forcing applied to the sheet enters the von K\'{a}rm\'{a}n
equations via the term $P$
or via boundary conditions at the
sheet's edges. We described the proper rescaling of $P$ in the last
section -- in this section we calculate the
rescaling of a particular kind of
in-plane forcing at the sheet boundaries.
Here and in the remainder of this work
we consider an external potential
which applies point forces to both vertices at the
ends of a ridge. Since
the applied forces have only delta function spatial extents
and are applied at points which remain stationary under ridge scaling,
we do not expect them to destroy the spatial scaling of the ridge
solution.
Therefore we may reasonably expect to find that the
equilibrium configuration of a ridge under a given compressive force
is identical to rescaled configurations
of ridges with different
material thickness and properly rescaled external force magnitudes. 


To calculate the proper rescaling 
of the forces on the vertices for a similarity
solution, 
we consider our forcing as a boundary condition consisting mainly
of an in-plane point force.
This force amounts to a
point stress at the edge of the sheet with the form
\begin{equation}
\sigma^{(o)}_{xx} = F_o \delta (y).
\end{equation}
So, to find similar scaled configurations of the sheet, we must scale 
$\sigma^{(o)}_{xx}$ the same way $\sigma_{xx}$ scales on the ridge. Since
$\gamma_{xx}$ scales as $\lambda^{2/3}$, 
\begin{equation}
\sigma_{xx} \sim Y h \gamma_{xx} 
\sim \kappa \lambda^{-4/3} X^{-2} 
\tilde{\gamma}_{xx},
\end{equation}
where $\tilde{\gamma}_{xx}$ is a dimensionless,
scale free function.
To express $\sigma^{(o)}_{xx}$ in similar fashion, we 
first substitute the
scale free $y$-variable $\tilde{y} = \lambda^{-1/3} y/X$, so that 
$\delta (y) = \lambda^{-1/3} X^{-1} \delta (\tilde{y})$. Thus, the 
proper scale free force can be written in terms of $F_o$ as
\begin{equation}
F_o = \frac{\kappa \lambda^{-1}}{X} \tilde{F}_o.
\label{eq:ftildeo}
\end{equation}

For reasons which will become clear in Section~\ref{ch:sims}, 
we cannot
measure the force applied to our ridge with very good
accuracy. However, we can measure
the inward displacement $\Delta$
of the ridge ends caused by this forcing. 
These two quantities may be 
related by assuming that the
work done by equivalent rescaled forces, 
given approximately by $F_o \Delta$,
scales the same as the total energy of the resting ridge
configuration. 
The total energy of a resting ridge scales 
as $\kappa \lambda^{-1/3}$, so equivalent values of $F_o \Delta/ \kappa$
will scale as $\lambda^{-1/3}$. Given the scaling of $\tilde{F}_o$
from Eq.\ref{eq:ftildeo}, the scale free
$\tilde{\Delta}$ must be related to the actual displacement by
\begin{equation}
\Delta = \lambda^{2/3} X \tilde{\Delta}.
\end{equation}
This scaling result has the predictable implication that the 
macroscopic strain $\Delta/X$ scales identically to $\gamma_{xx}$,
the local longitudinal strain on the ridge.

\section{Simulations}
\label{ch:sims}

\subsection{Numerics}

\begin{figure}

\center

\epsfig{file=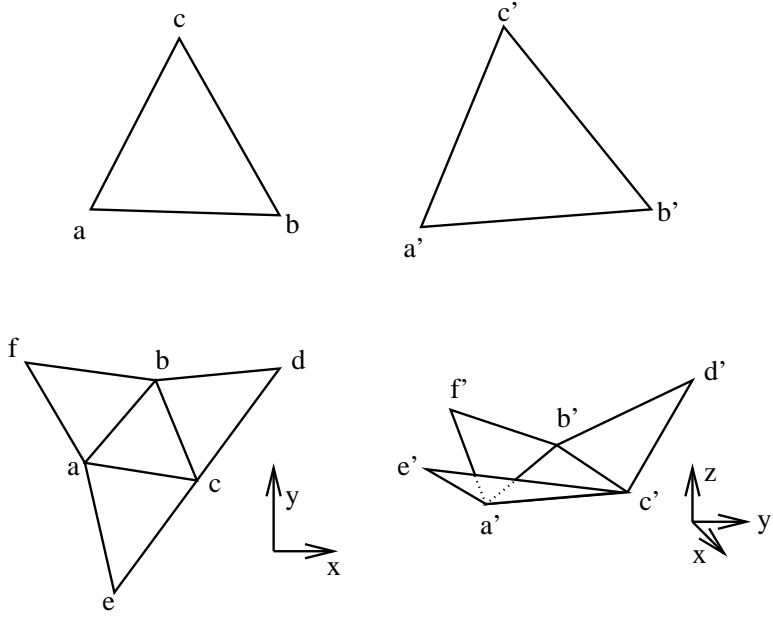, width=4in}

\caption{Finite elements for stretching and bending}
\label{fig:finiteelemnt}

\legend{The strain on each triangle is computed from the change in relative 
positions of its vertices. Curvature on one triangle 
is computed from the relative heights, normal to the triangle surface, 
of the vertices of the triangle plus the additional vertices of its 
nearest neighbor triangles.}

\end{figure}

\begin{figure}

\center

\epsfig{file=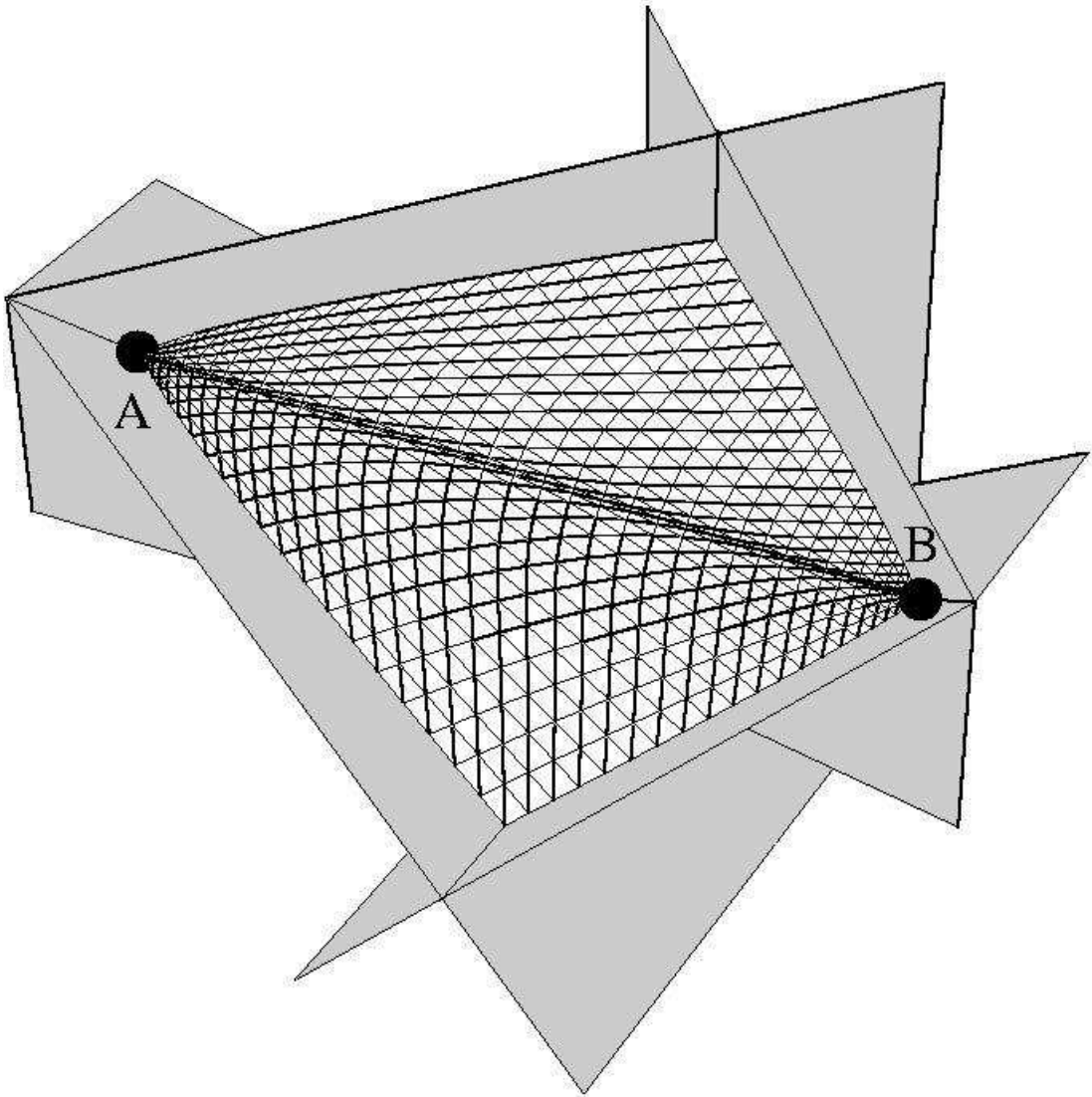, width=5.75in}

\caption{Simulational geometry}
\label{fig:simfig}

\legend{An equilibrium configuration of the simulated sheet 
(white grid) is
shown for a thickness aspect ratio $\lambda$ of $10^{-3}$. The dark
planes mark the location of the reflective planes to which the sheets
edges are confined. Points A and B are the center-points of the
repulsive $r^{-8}$ potential used to press on the vertices of the
simulated ridge.}

\end{figure}

We simulate an elastic sheet using a fixed triangular grid with variable
grid spacing (see Figure~\ref{fig:simfig}). 
Strains and curvature are taken to be constant across the
face of each triangle. Strain is calculated on each triangle by
measuring the relative deviation of its vertices from their predefined
strain free positions. Curvature is calculated on each triangle from the
relative heights normal to the triangle surface of 
the three triangles that share
sides with it (see Figure~\ref{fig:finiteelemnt}). 
The relative heights $z_i$ of the six points (a)~-~(f) are
fit to a function of the form
\begin{equation}
z_i = a_{1} + a_{2} x_i + a_{3} y_i + 
a_{4} \left( x_i \right)^2 +
a_{5} x_i y_i + a_{6} \left( y_i \right)^2, \ i=1,6
\label{eq:curvexpansion}
\end{equation}
where $x_i$ and $y_i$ are the material coordinates of the vertices.
Curvatures follow immediately from the
identification
\begin{gather}
C_{xx} = 2 \times a_{4}, \
 C_{xy} = a_{5},  \
C_{yy} = 2 \times a_{6}.
\label{eq:curvcoef}
\end{gather}
These six simultaneous equations for the $z_i$ in terms of 
$a_j$ are inverted at program initialization to save later computational
time.

The edges of the grid are constrained to lie in 
reflection planes of the minimal unit pictured in
Figure~\ref{fig:simfig}. 
For calculations of the curvature on triangles bordering these
planes, the triangles see mirror images of themselves across the planes.
The mutual attitudes of the reflection planes are such that the
sheet would meet them all at normal angles if it were perfectly flat
except for one sharp $90^\circ$ fold between two opposite corners. 
As shown in Figure~\ref{fig:grid}, the single ridge with this geometry 
is conceptually identical to one edge of a cube surface, provided the 
surface is constrained to be symmetric.
Two corners of the sheet
are the vertices of the ridge that forms along the fold-line when the
elastic energy is minimized.

The gridding was determined in two steps:
first the desired 
area was gridded
with an equilateral triangular lattice, then the 
positions of the grid points were remapped 
by the simultaneous transformation
\begin{equation}
x' = f(x, y), \ \
y' = g(x, y),  
\end{equation}
where $f$ and $g$ are fifth order polynomials~\cite{meshing} which are
constrained to be stationary on the edges and along the midline of the
grid. This mapping 
provided a fourth-order smooth gradient in grid spacings on the 
flat faces of the ridge while concentrating the
lattice spacing at the vertices by a factor of $10^3$ and across the
ridge-line by a factor of $10^2$ compared to the flat regions far from
the ridge. The
concentrations factors were chosen arbitrarily, within the limits of the
mapping, to make the gridding near the vertices as fine as possible,
since this is the region of largest
gradients in curvatures and strain. 
The gridding is visible in Figure~\ref{fig:simfig}.

Bending and stretching energies were assigned to
the curvature and strains
on each triangle using the forms for elastic energy
presented in~\cite{Seung.Nelson}
\begin{eqnarray}
e_{B} & = & \frac{1}{2} \kappa 
A \left( C_{ii} C_{jj} - \epsilon_{ik} \epsilon_{jl} 
C_{ij} C_{kl} \right) \\
e_{S} & = &  \frac{3}{16} G A 
\left( \gamma_{ii} \gamma_{jj} + 2 \gamma_{ij} \gamma_{ij} \right)
\end{eqnarray}
where $\epsilon_{ij}$ is the antisymmetric tensor,
$\kappa$ is a bending modulus, 
$G$ is the $2$-dimensional Young's
modulus, and $A$ is the area of the triangle. The stress
energy expression is
appropriate for a material with Poisson ratio of $1/3$. The 
coefficient of the Gaussian curvature energy
is not consistent with that 
given by Eq.~\ref{eq:kappaY} for a uniform elastic material, but
was chosen to maintain consistency with Lobkovsky et al.'s simulations 
in~\cite{science.paper,our.stuff,Alex}. The direct contribution of the 
Gaussian curvature to the bending energy is much less than that of the 
mean curvature. Separate simulations verified that
changing the value of 
the Gaussian curvature coefficient had no discernable 
effect in our data.
The physical thickness $h$ of the sheet is 
equal to $\frac{3}{4 \sqrt{2}} \sqrt{\kappa/G}$.

Pushing on the tips of the ridge is accomplished by introducing
repulsive potentials of the 
form $V(r) = C_P/\left| \vec{r}-\vec{x}_p \right|^8$ centered around 
two points, one on each line to which a vertex is constrained
(points A and B in Figure~\ref{fig:simfig}). The center
points are located at a distance $X$ from one another and symmetrically
placed with respect to the middle of the ridge.
These points lie where the vertices would be if the sheet 
were sharply creased -- relaxation of the ridge curvature
draws the vertices inward from these points for an unforced resting
ridge. The benefit of this potential is that it acts mainly on a
small but finite area around the vertex. In earlier simulations, 
simple pushing of the vertex itself led to local collapse of the vertex
tip without imparting any force on the main part of the ridge.
$C_P$ was varied to apply different loading. 

An inverse
gradient routine~\cite{conj_grad} was used to minimize the total
elastic and potential energy of the sheet as a function of the
coordinates of all the lattice points for given parameters 
$\kappa$, $Y$ and
$C_P$.

Using this routine
we found minimum energy configurations for ridges of aspect
ratio $\lambda$ ranging from $1.25 \times 10^{-3}$ to $1.77 \times
10^{-5}$. The upper bound on $\lambda$ was determined by the range of
validity of the ridge scaling solution -- 
above this value the width of the ridge becomes
comparable to that the sheet. At the other extreme,
for $\lambda < 10^{-5}$ the radius of
curvature at the ridge line becomes comparable to the spacing of our
lattice and the simulation ceases to be accurate. 

\subsection{Findings}
\label{sec:findings}

\begin{figure}[tbp]

\center

\epsfig{file=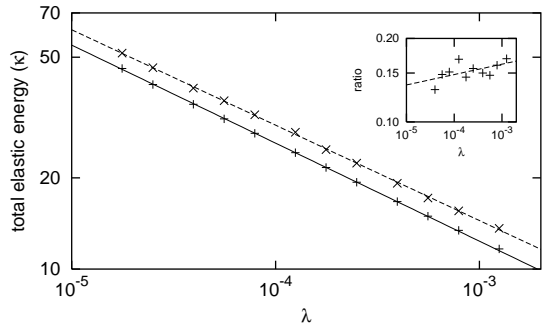}

\caption{Energy of ridges at rest and at the buckling threshold}

\legend{Straight lines are least squares fits to a scaling form 
$y=a x^b$.  In this plot 
$\lambda$ ranges from $1.25 \times 10^{-3}$ to $1.77 \times
10^{-5}$. The plot shows the total elastic energy
($E_B+E_S$) in the sheet after minimization.  
The scaling exponent fit for the resting
ridge values (lower line) was $-0.32$, 
the fit at the buckling threshold was $-0.31$. 
The inset shows the
difference between threshold energy and resting energy 
in units of the resting energy. 
This energy ratio is best fit by a scaling
exponent of $0.05 \pm 0.02$
and is consistent with a constant ratio.}

\label{fig:scaling}
\end{figure}

\begin{figure}[tbp]

\center

\scriptsize

\epsfig{file=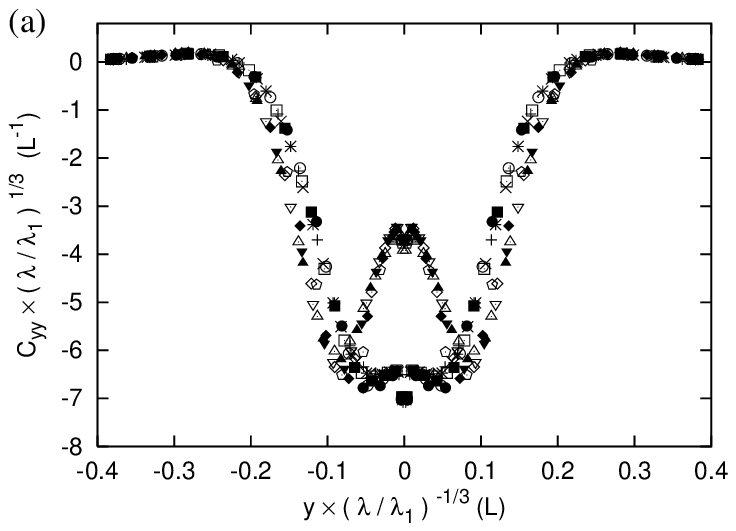}

\vspace{0.2in}

\epsfig{file=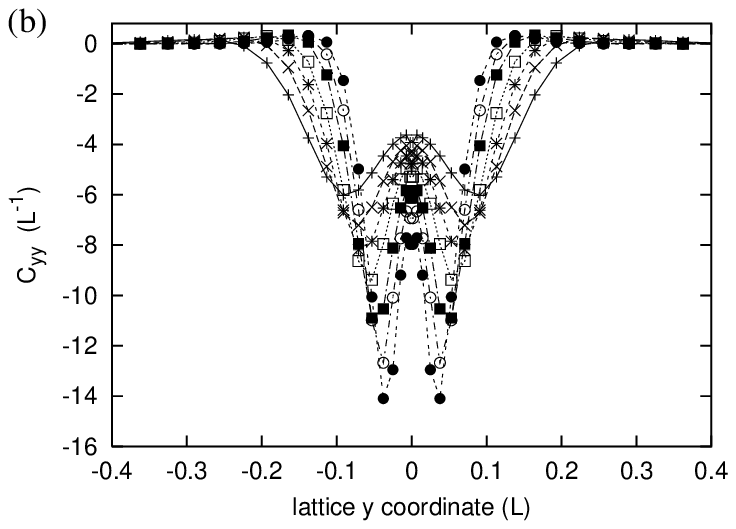}

\caption{Similarity solution for the ridge response
to forcing}

\legend{Both plots show $C_{yy}$, the curvature across the
ridge-line, versus the $y$ material coordinate on the line which bisects
the ridge-line. The data is for sheets with
seven different values of $\lambda$, ranging from 
$1.25 \times 10^{-3}$ to $1.25 \times 10^{-4}$. 
Plot (a) shows
$C_{yy} \times \left( \lambda / \lambda_1 \right)^{1/3}$ vs. 
$y \times \left( \lambda / \lambda_1 \right)^{-1/3}$ for 
the ridges at rest and for ridges with 
inward vertex displacement $\Delta(\lambda) = 
\Delta_1^{(c)} \times \left( \lambda / \lambda_1 \right)^{0.67}$,
where $\Delta$ is measured from the resting vertex
positions, $\lambda_1$ is the aspect ratio for the thickest sheet,
and  $\Delta_1^{(c)}$ was vertex displacement
at the buckling threshold for the
thickest sheet. The profiles with the large central peak are the
buckling threshold values.
(The small dimple 
in the data at $y=0$ is a numerical
artifact due to a discontinuity in the gridding density across the
ridge-line. For finer gridding this dimple goes away, while all other
local values of curvature remain constant.)
Plot (b) shows
unscaled  $C_{yy}$ versus $y$ for 
the buckling threshold profiles plotted in (a).}

\label{fig:rescaling}
\end{figure}

The plot in Figure~\ref{fig:scaling} shows
scaling of the total elastic energy
in the ridge versus $\lambda$ for ridges at rest ($C_{P}=0$) and at 
the buckling threshold. The data shown here is for ridges with 
dihedral angle $\frac{\pi}{2}$.
Scaling of the total elastic energy
for the resting configuration is
consistent with a $\kappa \lambda^{-1/3}$ dependence, in
agreement with prior theory and simulation~\cite{Alex,science.paper}.
Figure~\ref{fig:scaling} also shows that 
the elastic energy measured 
at the buckling threshold exhibits {\it exactly the same}
scaling as on resting ridges. 
This suggests that the ridge scaling developed for the resting ridge is
still applicable to the ridge with forces applied at its endpoints.
As we predicted in Section~\ref{sec:tipforcing}, this
particular form of forcing potential should not destroy
the length scaling of the ridge. 
The inset in Figure~\ref{fig:scaling} shows that that the energy
correction at the buckling threshold is nearly a constant fraction of
the total ridge elastic energy.

Scaling of the force response is
verified by the existence of a similarity solution for the ridge shape
as a function of tip displacement $\Delta$ (see Figure~\ref{fig:rescaling}).
We considered scaling of the equilibrium value of $C_{yy}$ 
along a line in the material coordinates which 
bisects the simulated ridge line.
As a consequence of the scaling exponents presented in 
Eq.~\ref{eq:scalings}, for an unforced ridge
plots of $C_{yy} \times \lambda^{1/3}$
versus $y \times \lambda^{-1/3}$ 
along this line should be independent of $\lambda$.
Extending this result to forced 
ridges, we found numerically that 
the rescaled cross-ridge curvature profiles were also
identical for forced ridges with the same equilibrium value of 
$\Delta \times \lambda^{0.67}$.
This $\Delta$ rescaling exponent is very close to the theoretical
value of $2/3$ derived above. 
Plot (a) in Figure~\ref{fig:rescaling} shows values of $C_{yy}$, 
along a line in the material coordinates which bisects the simulated
ridge line,
for several different sheet thicknesses and two different values
of rescaled ridge tip displacement $\Delta$. 
For comparison, the
unscaled $C_{yy}$ versus $y$ for a particular rescaled $\Delta$ is
shown in Figure~\ref{fig:rescaling}(b).

With our generic treatment of the
ridge force response we can rescale the observed 
configuration
at one thickness to that for the simultaneously
rescaled thicknesses and applied forces. 
It must be noted, however, that scaling of the
ridge response to forcing does not imply identical scaling of the
buckling threshold. Buckling of the ridge signals 
a bifurcation in the allowed equilibrium configurations at a critical
applied load~\cite{stoker,keller}. 
This is a completely separate topic, which we treat in
Section~\ref{ch:buckling}. 
There, we introduce a model for the buckling transition
which reproduces the observed scaling of the critical energy without
assuming it {\it a priori}.

\subsubsection{Dihedral angle scaling}
\label{app:simscaling}

In addition to the simulational data presented above for ridges 
with dihedral angle $\frac{\pi}{2}$, we also
simulated the force response up to the buckling point for ridges 
with different angles.
In this section, we present data for
ridges with thickness aspect ratios $\lambda$  
from $1.25 \times 10^{-3}$ to $2.5 \times 10^{-4}$ 
and with dihedral 
angles from $\frac{\pi}{2}$ to $\frac{7\pi}{10}$. In each of 
these simulations, the sheet had equal side lengths as before
and was held by 
reflective boundary conditions similar to those described above
to form a ridge between two corners.

\begin{figure}

\center

\epsfig{file=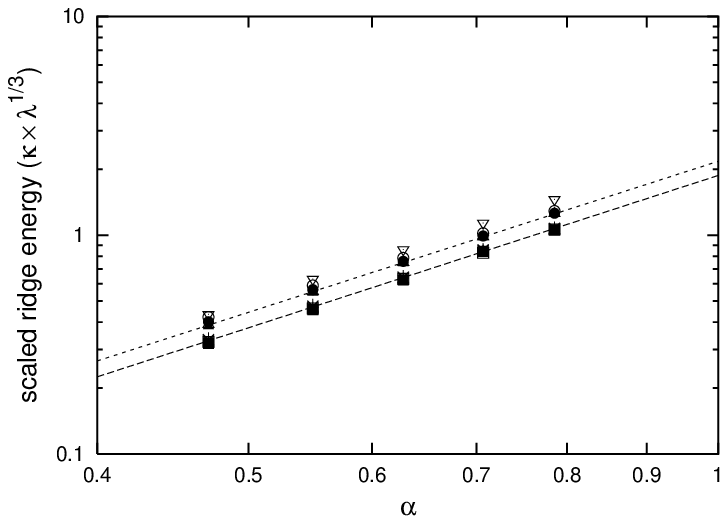}

\caption{Scaling of ridge energy with $\alpha$}
\label{fig:alphaens}

\legend{Data is shown for ridges with thickness aspect ratio $\lambda$
ranging from $1.25 \times 10^{-3}$ to $2.5 \times 10^{-4}$. 
The dihedral angles of the ridges are 
given by $\pi - 2 \alpha$. The energy
values for each thickness were rescaled by the predicted energy scaling
factor $\lambda^{1/3}$ so they all lie on a common line. Points on the
lower line were for ridges at rest and points on the upper line were for
ridges at the buckling threshold. The lines are fits to the data for
thickness $\lambda = 5.6 \times 10^{-4}$. The $\alpha$ scaling 
exponents for this thickness were $2.31$ for resting ridges and 
$2.29$ for ridges at their buckling threshold. The thinnest sheets
(``$\nabla$'' symbols) do show some deviation from the top line fit.}

\end{figure}

Lobkovsky showed~\cite{Alex} that for 
ridges with dihedral angle $\pi-2 \alpha$ the elastic
energy scales as $\kappa 
\alpha^{7/3}$. The best scaling fit to the 
resting energy of our ridges
had an exponent of $2.31$ (see
Figure~\ref{fig:alphaens}). 
Interestingly, for all dihedral angles and 
thicknesses studied, we found that the total elastic energy 
at the buckling threshold was always approximately $\universal$ greater 
than the resting ridge energy. 
The constancy of this ratio is discussed in 
Section~\ref{sec:diffangles}, once we have derived the 
buckling criterion.

\begin{figure}

\center

\epsfig{file=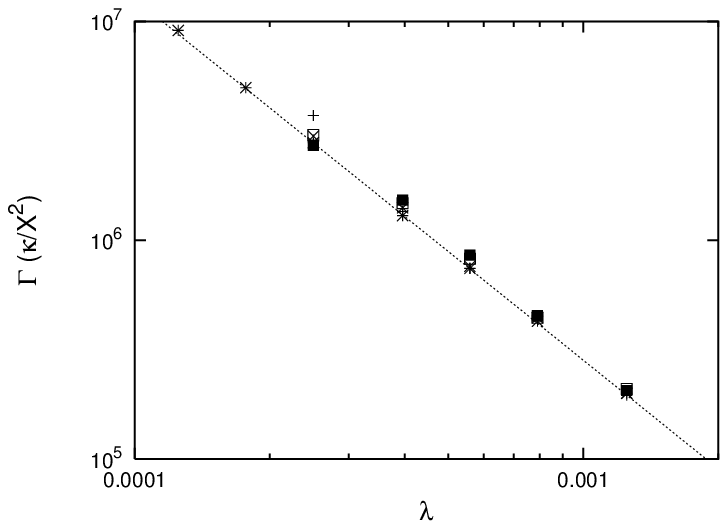}

\caption{Scaling of $\Gamma$ with $\alpha$}
\label{fig:alphagamma}

\legend{The compressibility modulus $\Gamma$ was calculated for ridges 
with thickness aspect ratios $\lambda$  
from $1.25 \times 10^{-3}$ to $1.25 \times 10^{-4}$ 
and with dihedral 
angles from $\frac{\pi}{2}$ to $\frac{7\pi}{10}$. The line is a scaling
fit for data with dihedral angle $\frac{\pi}{2}$ and has an exponent of
$-1.65$.} 
\end{figure}

We also found for all cases studied that the total energy in the ridge as a 
function of tip displacement $\Delta$ was well fit by the quadratic 
functional form
\begin{equation}
E = E_{o} + \frac{1}{2} \Gamma \left( \Delta - \Delta_{o} \right)^{2},
\end{equation}
where $E_{o}$ is the resting ridge energy and $\Delta_{o}$ is a 
numerically fit zero offset. Typical values of $\Delta_{o}$ were 
found to be within $10 \%$ of the resting ridge zero offset.
The value of $\Gamma$ was found to be nearly independent of dihedral 
angle, but was well fit by the scaling form 
$\Gamma \approx 3.2  \ \lambda^{-1.65} \ \kappa/X^{2}$ (see
Figure~\ref{fig:alphagamma}). For comparison, the
elastic energy of a thin strip of length $X$ and width $w$ 
whose ends are compressed inward by length $\Delta$
is approximately given by $E \approx \frac{1}{2} Yh
(\Delta/X)^2 X w$. If we take the width to be the ridge width 
$w = X \lambda^{1/3}$ and
substitute $Yh = \kappa/h^2 = \kappa \lambda^{-2}/X^2 $, 
then the energy becomes
$E \approx \frac{1}{2} \lambda^{-5/3} \kappa/X^2 \Delta^2$. 
Thus the compressibility of a thin 
flat strip with width on the order of
the ridge width is also of order 
$\lambda^{-5/3} \kappa/X^2$.

\section{The buckling transition}
\label{ch:buckling}

Detailed study of the buckling transition is complicated by an apparent
discontinuity between pre-buckled
and post-buckled states. When a sheet of paper or a tin can
is buckled new ridges appear suddenly, often accompanied by the 
popping sound of energy release~\cite{ericandalex}. 
In simulations we find that 
ridges buckle
at a repeatable value of inward tip displacement, but 
immediately after the buckling transition the minimal energy
configuration of
the sheet contains a fully formed new ridge. This is counter to our
intuition that the transition should be 
continuous~\cite{stoker,keller}, in which case the
immediately post-buckled state would be only infinitesimally different
than the pre-buckled. More importantly, our inability to observe 
intermediate stages in the growth of the buckled state
prevents us 
from directly seeing the shape of the assumed normal 
mode against which the ridge becomes unstable.

\begin{figure}

\center

\epsfig{file=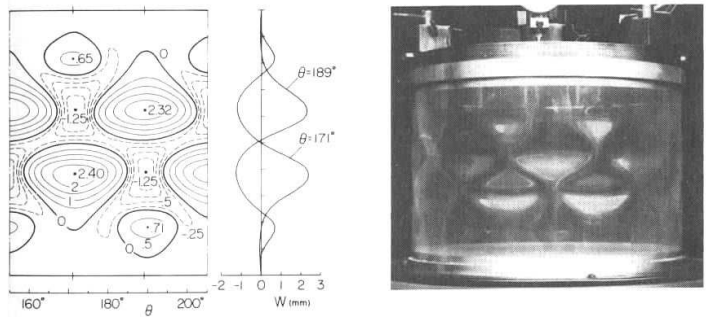, width=4in}

\caption{Diamond shaped buckling mode for a thin cylinder}
\label{fig:tincanmode}

\legend{This image was taken from~\cite{yamaki}. On the right is a 
photograph of an aluminum cylinder buckled by application of straight 
downward forces at its ends. On the left is a relief map of a 
simulated sheet with the same buckling pattern.}

\end{figure}

In this section we study the buckling transition in greater depth, through
more detailed analysis of the simulational results presented earlier. The
consequences of our analysis also lead us to run
further, more specialized simulations. Our aim is to tie the observed
behavior at the ridge 
buckling threshold to well known results concerning the
buckling of thin elastic cylinders. 

We begin by reviewing the salient
features of thin cylinder buckling. Under uniform axial compression
applied at its ends, a thin elastic cylinder will pass from a
state of uniform uniaxial curvature to an axially periodic diamond shaped
buckling pattern 
(see Figure~\ref{fig:tincanmode}).
Although the details of the geometry are different, we show that the
buckling of 
elastic cylinders or sections of a cylinder
is determined by elastic terms analogous
to those
which dominate the behavior of stretching ridges -- namely the
cross-ridge curvature and the ridge-line strain. We proceed to apply the
analysis developed for the cylinder 
to elastic ridges, finding that the
scaling it predicts for the buckling transition is consistent with our
simulational observations. We then do a normal mode analysis of the
buckling transition to determine the scaling of the lowest mode as it
approaches the point where the single ridge
configuration becomes unstable.
We numerically compute the lowest several normal modes for ridges under
various degrees of compression, and show the appearance of a mode with
swiftly decreasing eigenvalue which we believe accounts for the buckling
instability. Quantitative aspects of this mode's shape and its approach
to zero eigenvalue are discussed.

Finally, we argue that the apparent differences between the elastic
cylinder and the stretching ridge are inconsequential in regard to the
buckling instability. The main distinguishing trait of the stretching
ridge is that it maintains high internal stresses even when it is at
rest. We show that the critical load at which the ridge buckles is
determined by the ridge curvature and
the total strain along the length of the ridge, including the
pre-existing stress and the additional stress resulting from the applied
load. 
To demonstrate that the buckling transition only 
depends on two parameters, the transverse curvature and the total 
longitudinal stress, we present data from several additional
simulations with different geometries. Though these new simulated sheets 
buckle at different total stresses and curvatures, they all buckle at 
the same ratio of these two quantities. We also discuss an observed 
universality of the additional energy required to break a ridge.

As a last note, we address the jump in position and energy at the 
buckling transition. Even though the appearance of unstable modes for 
cylinders is the result of a continuous bifurcation in phase space, 
cylinders also jump discontinuously in energy upon 
buckling. We justify this jump in terms of the non-linear growth of 
the buckling mode. We also argue that the final wavelength of the 
buckling pattern on a ridge need not be the wavelength of the 
instability.

\subsection{Stability of thin elastic cylinders}
\label{sec:cylinders}

Our treatment of the stability of thin elastic cylinders and sections 
of a cylinder under a
compressive load mainly follows that presented
in~\cite{yamaki}. To avoid additional boundary conditions,
we make our argument for a complete cylinder -- however,
the stability condition we find is local, so it can be directly 
applied to an angular section of a cylinder with the same local stress 
and curvature fields. We discuss local buckling and angular shell 
sections at the end of this section. 

We consider an elastic cylinder of thickness $h$,
length $L$, radius $R$ and Young's modulus $Y$. To assess the stability
of the cylinder under a compressive force $F=2 \pi R \sigma$
uniformly applied along its edges, we consider the stability of the
force \vonkarman equation, Eq.~\ref{eq:forcevk}, against infinitesimal
displacements. The force $F$ is conveniently expressed so that the
resulting longitudinal stress in the unbuckled cylinder is $\sigma$.
Under the application of the force at its ends, the cylinder will
naturally undergo some compression along its length, which will in turn
cause it to expand radially. We are not concerned with these
distortions, but only consider them as the equilibrium solution to the
\vonkarman equations to which we add an infinitesimal displacement
which will grow into a buckled solution. 
We define a local coordinate system everywhere 
on the cylinder with $x$
direction along its length, $y$ azimuthal, and $z$ normal to the
surface. Infinitesimal displacements in these three directions are
labeled $u$, $v$, and $w$ respectively. In this
frame, $\hat{z}$ points radially inwards.

\begin{figure}

\center

\epsfig{file=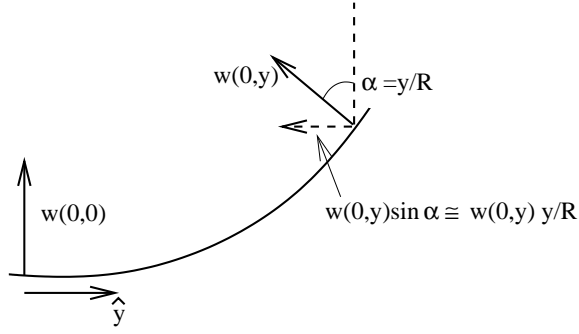, width=3in}

\caption{Coordinates in cylinder frame}
\label{fig:cylframe}
\end{figure}

Because our cylinder already has a curvature field $C_{yy} = 1/R$, the
relation between the additional displacements $u$, $v$, and $w$ and the
resulting strain and curvature fields is not as simple as for  flat
sheets. Up to an additive constant, the local embedding of the cylinder
into $\calR^3$ is given by:
\begin{equation}
\vec{r}(\vec{x}) = \left( x+u, (y+v)(1-w/R), w + 
\frac{1}{2}\frac{(y+v)^2}{R} \right).
\end{equation}
This expression accounts for the rotation of our local frame, as
demonstrated in Figure~\ref{fig:cylframe}. 
Referring to Equations~\ref{eq:straindef}
and~\ref{eq:curvdef}, the expression for the {\it additional} strain and
curvature due to our infinitesimal displacements are, to first order 
in $u$, $v$ and $w$,
\begin{gather}
\gamma'_{xx} = \frac{\partial u}{\partial x}, \
\gamma'_{yy} = \frac{\partial v}{\partial y} - \frac{w}{R}, \
\gamma'_{xy} = \frac{1}{2} \left( \frac{\partial u}{\partial y} + 
\frac{\partial v}{\partial x} \right)
\label{eq:newstraindispl}
\end{gather}
\begin{gather}
C'_{xx} = \frac{\partial^2 w}{\partial x^2}, \ 
C'_{yy} = \frac{\partial^2 w}{\partial y^2} + \frac{1}{R} 
\frac{\partial v}{\partial y}, \ 
C'_{xy} = \frac{\partial^2 w}{\partial x \partial y} + \frac{1}{R} 
\frac{\partial v}{\partial x}.
\label{eq:curveonr}
\end{gather}
Here the primes denote the infinitesimal corrections to the
equilibrium fields. For equilibrium small displacements, the
terms in Eq.~\ref{eq:curveonr}
involving derivatives of $v$ are typically
much smaller than those involving derivatives of $w$, so they are
neglected in the following treatment. The solutions we find for 
$u$, $v$ and $w$ are consistent with this approximation for the
values of
$h/R$ considered.

Empirically, thin elastic
cylinders typically buckle in a diamond pattern like that shown in
Figure~\ref{fig:tincanmode}. The longitudinal and azimuthal periodicities of 
these patterns vary. To test the stability of the shell against these
buckling modes, we consider a family of
infinitesimal displacement of the form:
\begin{gather}
u = A e^{i r x/R} \cos ( N y/R ), \nonumber\\
v = B e^{i r x/R} \sin ( N y/R ), \nonumber\\
w = C e^{i r x/R} \cos ( N y/R ),
\label{eq:bucklemode}
\end{gather}
where the periodicities $r$ and $N$ are free variables.
This form of the buckling mode neglects boundary effects, and so it is 
most accurate for very long cylinder, for which $R/r \gg L$.

We neglect higher order corrections 
due to the geometric \vonkarman equation in our consideration
of infinitesimal
displacements, but the force \vonkarman equation is 
not valid unless
the additional constraint of in-plane force equilibrium
is satisfied. From Eq.~\ref{eq:eqstrain}, this requires 
$\partial_i \sigma'_{ij} = 0$. 
Using the relation of stress to strain from
Eq.~\ref{eq:stressstrain} combined with Eq.~\ref{eq:newstraindispl},
this yields two equations:
\begin{gather}
\partial^2_x u + \frac{1-\nu}{2} \partial^2_y u + \frac{1+\nu}{2}
\partial_x \partial_y v - \frac{\nu}{R} \partial_x w = 0
\nonumber, \\
\nonumber \\
\frac{1+\nu}{2} \partial_x \partial_y u + \frac{1-\nu}{2} \partial^2_x v
+ \partial^2_y v - \frac{1}{R} \partial_y w = 0.
\end{gather}
These equations can be used to solve for $A$ and $B$ in terms of $C$,
yielding:
\begin{equation}
\frac{A}{C} = -ir \frac{\nu r^2 - N^2}{(r^2 + N^2)^2}, \ \ 
\frac{B}{C} = N \frac{(2+\nu) r^2 + N^2}{(r^2 + N^2)^2}.
\label{eq:ABC}
\end{equation}

The equilibrium cylinder configuration becomes unstable to the
combined
displacements in Eq.\ref{eq:bucklemode} when the resulting normal
force, calculated from the force \vonkarman equation, goes to zero.
To linear order in the small displacements,
this condition will be met when
the terms linear in $A$, $B$ and $C$ satisfy
\begin{equation}
\kappa \nabla^2 C'_{ii} = \frac{1}{R} \sigma'_{yy} + \sigma C'_{xx}.
\label{eq:unstablevk}
\end{equation}
Using Eq.~\ref{eq:stressstrain} to express $\sigma'_{yy}$
in terms of $\gamma'_{yy}$ and $\gamma'_{xx}$, the above equation can be
written in terms of the small displacements as:
\begin{equation}
\kappa \nabla^2 ( \partial^2_x w + \partial^2_y w ) + 
\sigma \partial^2_x w - \frac{12 \kappa }{R h^2} 
\left( \partial_y v + \nu \partial_x u - \frac{w}{R} \right) = 0,
\label{eq:bucklebifurcation}
\end{equation}
where we have used Eq.~\ref{eq:kappaY} to express all elastic moduli
in terms of $\kappa$.

Substituting the buckling form of Eq.~\ref{eq:bucklemode} 
directly in Eq.~\ref{eq:bucklebifurcation} and using 
Eq.~\ref{eq:ABC} to eliminate $A$, $B$, and $C$ yields the
bifurcation condition
\begin{equation}
\frac{12 (1-\nu^2) R^2 \kappa}{h^2} r^4 + (r^2 + N^2)^2
\left( \kappa (r^2 + N^2)^2 - \sigma R^2 r^2 \right) = 0.
\label{eq:bifcond}
\end{equation} 
If we define 
\begin{equation}
\eta = \frac{r^2}{(r^2 + N^2)^2},
\label{eq:etadef} 
\end{equation}
we can write Eq.~\ref{eq:bifcond} as
\begin{equation}
\frac{12 (1-\nu^2) R^2 \kappa }{h^2} \eta^2 - \sigma R^2 \eta + \kappa = 0.
\label{eq:stableeta} 
\end{equation} 
Solving for $\sigma$ yields
\begin{equation}
\sigma = \frac{\left( 12 (1-\nu^2) 
\left(\frac{R}{h}\right)^{2} \eta^{2} 
+ 1 \right) \kappa}{R^{2} \eta} .
\label{eq:solveeta}
\end{equation} 
This function has a minimum in $\eta$ when
\begin{equation}
\eta \equiv \eta_{cl} = \left( 12 (1-\nu^2) \right)^{-1/2} 
\frac{h}{R} .
\label{eq:etaclassic}
\end{equation}
Where $\eta_{cl}$ is called the ``classical'' buckling 
value of $\eta$. The corresponding minimum value of $\sigma$ is 
denoted as $\sigma_{cl}$:
\begin{equation}
\sigma_{cl} = 4 \sqrt{3 (1-\nu^2)} \ \frac{\kappa}{R h} .
\label{eq:cylbreakstress}
\end{equation}
This is referred to as the ``classical'' value
of the buckling stress~\cite{yamaki,timoshenko}. 
The classical stress is the smallest applied load under which
the cylinder can buckle, provided that a 
solution for the corresponding value of
$\eta$ is allowed.\footnote{It has been found
experimentally that, depending on boundary conditions, initial 
imperfections and deviations 
from the diamond pattern near the cylinder ends 
can lower the bucking stress by up to $50\%$~\cite{yamaki,timoshenko}.
However, our treatment is sufficiently accurate to establish the 
length and energy scale at which cylinders will buckle when boundary 
effects are neglected.}

Our limit of a thin sheet corresponds to the limit
$\eta_{cl}\rightarrow 0$. The variables $r$ and $N$ are not uniquely 
determined by this stability
treatment, beyond the requirement that they satisfy 
Eq.~\ref{eq:etadef}. However, for each set of $r$ and $N$ there is a 
unique value of $\eta$ and therefore of
$\sigma \ge \sigma_{cl}$, determined from Eq.~\ref{eq:solveeta}, at 
which the cylinder becomes unstable to buckling with that mode.
If $r$ and $N$ are taken to be continuum 
variables, then there is an entire family of solutions which 
satisfy $\eta = \eta_{cl}$. In real cylinders the 
preferred buckling wavenumbers are
determined by boundary conditions and initial 
imperfections. Azimuthal periodicity and the 
requirement that the wavelength be
commensurate with the finite length of real cylinders severely
limits the number of allowed solutions for cylinders which are 
relatively thick. In such cases,  
there may only be a handful of allowed
combinations of $r$ and $N$ for which $\eta \approx \eta_{cl}$, 
with corresponding
threshold instability value of $\sigma$
close to $\sigma_{cl}$.
Also, it has been shown~\cite{timoshenko} that 
initial curvature imperfections in the cylinder can 
break the degeneracy in $r$ and $N$.

\begin{figure}

\center

\epsfig{file=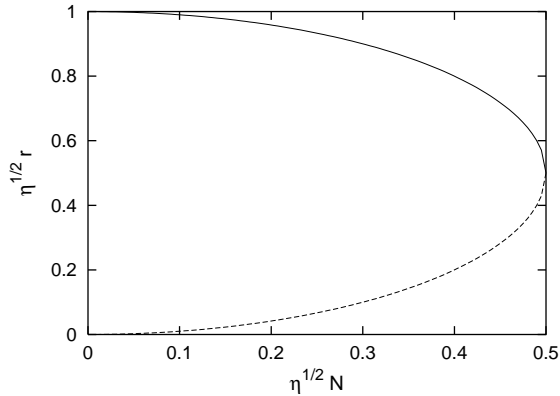}

\caption{Relation between $r$, $N$, and $\eta$}
\label{fig:etarange}

\legend{The solid line is the higher root
of Eq.~\ref{eq:rN}, while the
dashed line is the lower root.}

\end{figure}

Without yet placing any constraints on the allowed 
values of $r$ and $N$, we explore the limits imposed by 
Eqs.~\ref{eq:etadef}~and~\ref{eq:etaclassic}. 
Solving Eq.~\ref{eq:etadef}
for the wavenumber $r$ in terms of $N$ and $\eta$
gives
\begin{equation}
r \sqrt{\eta} = \frac{1}{2} \left[ 1
\pm \left( 1 - 4N^2 \eta \right)^{1/2}
\right].
\label{eq:rN}
\end{equation}
Thus $r$ can take values between zero and $\eta^{-1/2}$
and $N$ can range from zero to $\frac{1}{2} \eta^{-1/2}$.
Possible combinations of $r \sqrt{\eta}$ and 
$N \sqrt{\eta}$ are shown in Figure~\ref{fig:etarange}.
For most allowed combinations of $r$ and $N$, 
$r$ is on the order of 
$\eta^{-1/2}$. 
For small $r$ and $N \eta^{1/2}$ 
we can expand the lower root of Eq.~\ref{eq:rN},
finding $r \approx \eta^{1/2} N^2$. 
Substituting the value of $\eta_{cl}$ 
from Eq.~\ref{eq:etaclassic} gives
two limits for large $R/h$:
\begin{equation}
\sqrt{R/h} \ge r \ge N^2 \sqrt{h/R},
\label{eq:tworroots}
\end{equation}
where for the lower 
limit $N \ll \sqrt{R/h}$. 
Implications of both bounds for ridge buckling will be
described in later sections.


For comparison with later results, we note that the classical
breaking stress
given in Eq.~\ref{eq:cylbreakstress}
implies a breaking strain of order
\begin{equation}
\gamma_{xx} = \left(3(1-\nu^2)\right)^{-1/2} \ \frac{h}{R} 
\approx 0.61 \ \frac{h}{R},
\label{eq:cylbreakstrain}
\end{equation}
where we have assumed $\sigma_{yy} \approx 0$
and have used $\nu=1/3$ to correspond with our simulations.
Thus, the total energy input required to buckle the cylinder is
\begin{equation}
\frac{1}{2} 
\int \gamma_{xx} \sigma dA = 4 \pi (1-\nu^2) \frac{\kappa L}{R},
\end{equation}
which is on the same order as the energy required to bend the
cylinder out of a flat sheet
\begin{equation}
\frac{1}{2} 
\int \kappa C_{yy}^2 dA = \pi \frac{\kappa L}{R}.
\end{equation}

As we mentioned above, the threshold condition for cylinder buckling 
can also be considered as a local condition on an angular
section of a cylinder. 
The form of the buckling displacements which we posited in
Eq.~\ref{eq:bucklemode} is a global motion, but we can construct a 
localized wave packet from combinations of these modes. Especially for 
short wavelength modes, there are many different modes with 
corresponding $\eta$ near $\eta_{cl}$, so a localized packet can be 
constructed with critical stress very close to $\sigma_{cl}$. 
The only requirement for this to work 
is that the width of the packet be 
at least a couple times longer than the principal buckling 
wavelength. From Eq.~\ref{eq:tworroots}, the smallest packet must 
therefore have a width of at least $2\pi R/r = 2\pi \sqrt{hR}$.
All of the buckling motion may be 
localized within such a self-contained
wave packet, so the buckling threshold condition should 
be determined only by aspects of the
cylinder defined within the wave packet. 
In this instance, it is appropriate to 
rewrite Eq.~\ref{eq:cylbreakstrain} with the local
curvature $C_{yy}$ in place of $1/R$, so the local buckling condition
for $\nu = 1/3$ is:
\begin{equation}
\gamma_{xx} / C_{yy} = 0.61 h.
\label{eq:cylbreakratio}
\end{equation}
Thus, if the ratio of strain to curvature
locally surpasses the threshold 
value in Eq.~\ref{eq:cylbreakratio}
over a region of spatial 
extent greater than $2\pi \sqrt{hR}$, local buckling can occur just 
at that point. Experimentally, cylinders are often observed to begin 
buckling locally instead of all at once, due to inhomogeneities in the 
cylinder material and uneven forcing~\cite{timoshenko}.

\subsection{Application to ridge stability}
\label{sec:ridgestability}

It is immediately apparent that the elastic terms which dominate
the buckling behavior of cylinders are analogous to
those which determine the scaling behavior of ridges. The largest 
terms in the stability condition, Eq.~\ref{eq:unstablevk},
are those proportional to the transverse
curvature $1/R$ and the longitudinal strain $\sigma$. 
Likewise for the ridge, the balance between 
transverse curvature and longitudinal strain 
determine the ridge's shape and energetics both at rest and under
compression. Furthermore, if we substitute the scaling form of the ridge
curvature $1/R = C_{yy} \sim \lambda^{-1/3}/X $ into 
Eq.~\ref{eq:cylbreakstrain} for the critical strain of the cylinder
under applied load, we find $\gamma_{cr} \sim \lambda^{2/3}$, which is
exactly the scaling we observed for the ridge in
Section~\ref{sec:findings}. 

We therefore anticipate that the buckling mode
of the ridge should be roughly the same as that of a cylinder of 
material thickness $h$, length
$L=X$, and radius $R=\lambda^{1/3} X$, where X is the ridge length and
$\lambda$ is the thickness aspect ratio defined in
Sec.~\ref{sec:vonkarman}. We
picture the boundary layer of the ridge as behaving like an angular
section of a cylinder, with some semi-rigid boundary conditions at the
edges where the boundary layer meets the ridge flanks. This
picture makes two assumptions: first that the real buckling mode of the
ridge is localized on the boundary layer
and second that the longitudinal curvature on
the ridge-line doesn't change the critical strain scaling.
The first assumption is supported by numerical evidence in the next
section. Also, the ridge-line is observed to absorb nearly all
the stress of our applied load without noticeably 
changing the shape of the ridge flanks, so it is plausible that the
boundary layer only sees the rest of the sheet as a set of boundary
conditions. 

In the remainder of this section, we calculate the correction to the
buckling stress resulting from the non-zero longitudinal curvature.
For ridges with typical aspect ratio 
$\lambda=10^{-3}$ or less we anticipate that 
the longitudinal curvature is too small to have
a pronounced effect on the buckling transition. 
We argued earlier that the longitudinal radius of 
curvature goes to zero as $\lambda^{1/3} X$, so for an aspect ratio of 
$\lambda=10^{-3}$ this
radius of curvature already is of order $10X$. The buckling wavelength
couldn't be longer than the ridge, so the buckling
deformational mode should not
be strongly affected by the smaller curvature.

\begin{figure}

\center
\epsfig{file=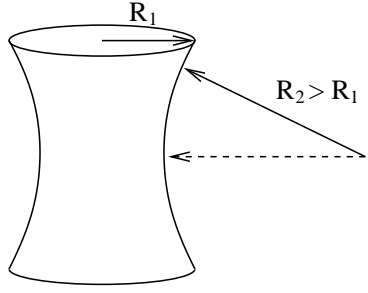, width=2in}

\caption{Distorted cylinder with one large and small curvature}
\label{fig:twocurv}
\end{figure}

In order to treat the longitudinal curvature rigorously, we repeat the
derivation of the last section for a surface with curvature $R_1$ in the
$y$ material direction and $-R_2$ in the $x$ direction. We take 
$R_2 \gg R_1$. These curvature fields cannot globally
describe a real surface,
but for $R_2$ small, we may picture a surface such as that in 
Figure~\ref{fig:twocurv}, where 
$R_1$ is nearly constant over the length of the object.
For such a surface, the strain-displacement relations analogous to
Eq.~\ref{eq:newstraindispl} become
\begin{gather}
\gamma'_{xx} = \frac{\partial u}{\partial x} + \frac{w}{R_2}, \
\gamma'_{yy} = \frac{\partial v}{\partial y} - \frac{w}{R_1}, \
\gamma'_{xy} = \frac{1}{2} \left( \frac{\partial u}{\partial y} + 
\frac{\partial v}{\partial x} \right)
\label{eq:twocurvstraindispl}
\end{gather}
and the in-plane strain equilibrium equations become
\begin{gather}
\partial^2_x u + \frac{1-\nu}{2} \partial^2_y u + \frac{1+\nu}{2}
\partial_x \partial_y v - \frac{\nu}{R_1} \partial_x w +
\frac{1}{R_2} \partial_x w = 0
\nonumber, \\
\frac{1+\nu}{2} \partial_x \partial_y u + \frac{1-\nu}{2} \partial^2_x v
+ \partial^2_y v - \frac{1}{R_1} \partial_y w +
\frac{\nu}{R_2} \partial_y w = 0.
\end{gather}

If we substitute the buckling mode from Eq.~\ref{eq:bucklemode} into the
above equations, the ratios of buckling coefficients become
\begin{gather}
\frac{A}{C} = \frac{-ir }{(r^2 + N^2)^2} \left( \nu r^2 - N^2 - 
\frac{R_1}{R_2} \left( (2+\nu) N^2 + r^2 \right) \right)
\nonumber , \\ 
\frac{B}{C} = \frac{N }{(r^2 + N^2)^2} \left( (2+\nu) r^2 + N^2 - 
\frac{R_1}{R_2} \left( \nu N^2 - r^2 \right) \right).
\label{eq:ABC2}
\end{gather}

Eq.~\ref{eq:bucklebifurcation}
for force balance of the infinitesimal displacement
gains several additional terms
from the new curvature, 
\begin{multline}
\kappa \nabla^4 w + 
\sigma \partial^2_x w - \frac{12 \kappa }{R_1 h^2} 
\left( \partial_y v + \nu \partial_x u - \frac{w}{R_1}
+ \nu \frac{w}{R_2} \right) \\
+ \frac{12 \kappa }{R_2 h^2} 
\left( \partial_x u + \nu \partial_y v - \nu \frac{w}{R_1} 
+ \frac{w}{R_2} \right)= 0.
\label{eq:bucklebifurcation2}
\end{multline}
Solving as before yields the equation
\begin{equation}
\frac{12 (1-\nu^2) R_1^4 \kappa}{h^2} 
\left( \frac{r^2}{R_1} - \frac{N^2}{R_2} \right)^2
+ (r^2 + N^2)^2
\left( \kappa (r^2 + N^2)^2 - \sigma R_1^2 r^2 \right) = 0.
\end{equation} 
If we define 
\begin{equation}
\eta' = \frac{r^2 - \frac{R_1}{R_2} N^2}{(r^2 + N^2)^2}, \ \
\sigma' = \frac{r^2 \sigma}{r^2 - \frac{R_1}{R_2} N^2},
\end{equation}
We can proceed as before to find the lowest allowed value for 
$\sigma'$ as a function of $\eta'$. The corresponding value
for $\sigma$ is
\begin{equation}
\sigma = 4 \sqrt{3 (1-\nu^2)} \ 
\frac{r^2 - \frac{R_1}{R_2} N^2}{r^2}
\frac{\kappa}{R h} .
\label{eq:ridgebreakstress}
\end{equation}
This stress is {\it lower} than the breaking stress for a cylinder, but
will approach the same value for $\frac{R_1}{R_2} \rightarrow 0$.
This justifies our above assumption that the minor $C_{xx}$ curvature 
has a weak effect on the buckling threshold.

\subsection{Numerical investigation of buckling modes}

\begin{figure}

\center

\epsfig{file=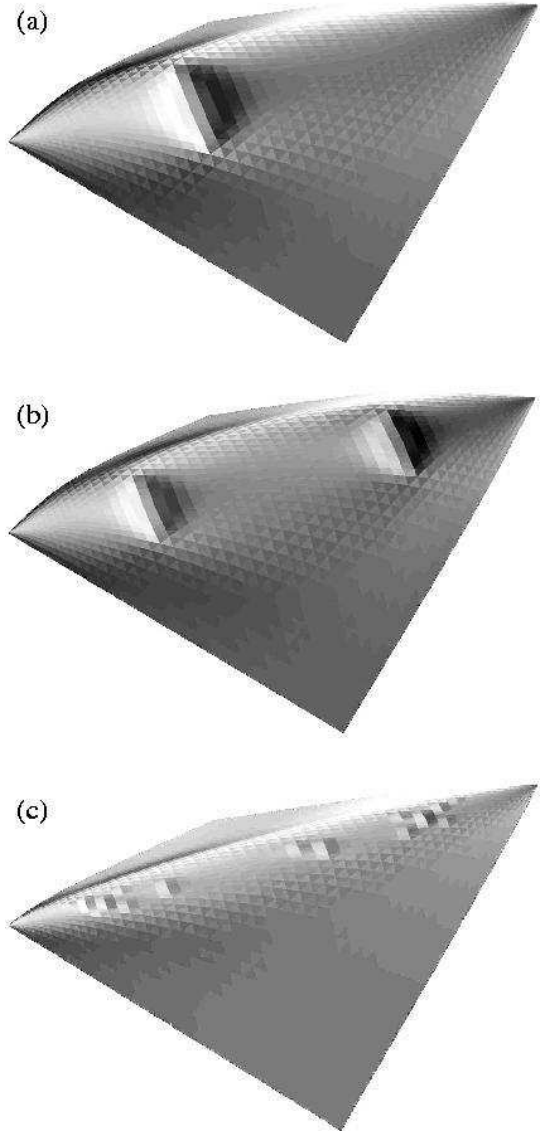, width=2.8in}

\caption{Observed post-buckling configurations}
\label{fig:buckled}

\legend{Images (a) and (b) show the buckling pattern for ridges with 
thickness aspect ratio $\lambda=1.25 \times 10^{-3}$. The 
configuration shown in (a) was at the smallest step past the buckling 
threshold that we simulated for this thickness. Image (b) shows a 
ridge with tip displacement further past the threshold value. Image 
(c) shows a buckled configuration for a ridge with thickness aspect 
ratio $\lambda=2 \times 10^{-4}$. Lighting and 
shading were chosen to emphasize physical features.}

\end{figure}

As mentioned above, our initial observations of the buckling 
transition came from detailed simulations in which we approached the 
buckling threshold with very small steps in inward vertex 
displacement. In all cases, the buckling transition was accompanied 
by a downward jump in total ridge energy and a large decrease in the 
strain along the ridge. 
There was no noticeable change in the vertex 
position after buckling. This assures us that the sudden jump is not 
aided by any work done on the ridge due to the springiness of our 
potential. 

The transition to the buckled state was of course accompanied by the 
appearance of additional ridges and vertices in our simulated 
sheets. Since the new ridges and vertices appeared in regions which 
were not finely gridded on our lattice they were often 
accompanied by curvatures which were large on the local scale of 
inverse lattice spacing. For this reason we 
do not claim that the buckling patterns we 
observed match in detail the real buckling pattern of a physical 
sheet, though they should be qualitatively correct.
The types of buckling 
pattern we observed are illustrated in Figure~\ref{fig:buckled}. For 
relatively thick sheets, the initial buckled state was like that in 
Figure~\ref{fig:buckled}(a) -- one new ridge appeared
across the original ridge, with length on the order of the unbuckled 
ridge width, positioned about two fifths of the way along the 
original ridge. In simulations with the same sheet thickness but 
larger steps in hard-wall potential position (so that the initial step 
across the buckling threshold pushed deeper into the 
buckled state), the first observed buckled state was like that in
Figure~\ref{fig:buckled}(b), with two new ridges positioned 
symmetrically about the midpoint of the original ridge. For 
thinner sheets like that shown in Figure~\ref{fig:buckled}(c) 
the buckling pattern consisted of a number of smaller ridges.
As with the larger ridges in (a)~and~(b),
these smaller ridges had length on order of the 
original ridge width and were clustered near locations 
about two fifths of the way along the original ridge. This is consistent
with the buckling patterns on a tetrahedron seen in~\cite{our.stuff}. In
that system, which is closely related to our own, buckling was
accompanied by the appearance of two large transverse ridges, which were
symmetrically spaced about one quarter of the way along the original
ridge. In~\cite{our.stuff}, only highly buckled states of the
tetrahedron were observed, so we cannot
be sure of where the new ridges initially 
formed on the tetrahedron (we have observed our own ridges to
change position slightly with the growth of the buckled state). 
This issue is revisited in
Section~\ref{sec:longerflanks}.

\subsubsection{Normal modes}

Since there seems to be some randomness involved in the selection of 
the postbuckled state accessible to our simulations, and since 
directly after buckling the system passes through non-equilibrium
states of intermediate energy which we cannot directly observe,
we desire to learn what we can about the buckling mode before 
the ridge buckles. Prior to buckling, the infinitesimal displacements 
given in Eq.~\ref{eq:bucklemode} 
would result in a restoring force normal to 
the surface opposing the growth of buckling mode. As the stress 
$\sigma$ approaches the buckling stress $\sigma_{cl}$, this restoring 
force goes to zero. Because the boundary conditions of a cylinder or 
a ridge enforce selection rules on the allowed buckling wavelengths, the 
buckling mode must be part of a discrete spectrum of eigenmodes for 
motion of the sheet. Very near the transition, the buckling mode is 
very soft, and at some point before the associated eigenvalue goes to 
zero it must have the lowest normal mode eigenvalue. 

Ignoring the small corrections calculated 
in Section~\ref{sec:ridgestability},
we can surmise how quickly the buckling mode approaches zero eigenvalue
with applied stress by expanding Eq.~\ref{eq:bucklebifurcation}
around $\sigma = (\sigma_{cl} - \sigma_{\delta})$.
For non-zero $\sigma_{\delta}$, the left hand side of
Eq.~\ref{eq:bucklebifurcation} will not equal zero, but will instead 
equal the restoring force per unit area linear in the buckling mode
amplitude, which we denote by $P_{\delta}$.
If we also ignore 
the small changes in transverse curvature $1/R$ with the applied 
stress, then the $\sigma_{cl}$ term cancels all the terms on the left 
hand side of Eq.~\ref{eq:bucklebifurcation}
except the $\sigma_{\delta}$ term, and we 
are left with
\begin{equation}
P_{\delta} = \sigma_{\delta} \partial^2_x w = 
\frac{-C \sigma_{\delta} r^{2}}{R^{2}}  e^{i r x/R} \cos ( N y/R ) ,
\end{equation}
where we have substituted the form for the buckling mode from 
Eq.~\ref{eq:bucklemode}. The work required to cause this displacement 
is then
\begin{equation}
W_{\delta} = \int P_{\delta} w \approx 
\left| P_{\delta} w \right| R X,
\end{equation}
where $R$ is the ridge radius of curvature, $X$ is its length,
and we again assume that the buckling motion is confined to the 
ridge boundary layer. The resulting expression is quadratic in the 
displacement amplitude $C$, and so by our previous ansatz we can 
identify it with the spring constant of the 
eigenmode which leads to buckling, by
$W_{\delta} 
\equiv \frac{1}{2} K_{\delta} C^{2}$. Expressing $\sigma_{\delta}$
in units of $\sigma_{cl}$, we can write
\begin{gather}
K_{\delta} \approx
\frac{ \kappa}{R^{2}} \frac{X}{h} \tilde{\sigma}_{\delta} r^{2} \\
\tilde{\sigma}_{\delta} \equiv \sigma_{\delta}/\sigma_{cl}
\end{gather}
Finally, if we substitute in the scaling form of the ridge curvature for
$1/R$ and the limiting
scalings of the classical buckling value of wavevector $r$, 
the expected spring constant of the buckling mode becomes
\begin{equation}
K_{\delta} \approx 
\begin{cases}
\frac{ \kappa}{X^2} \lambda^{-7/3} 
\tilde{\sigma}_{\delta} & 
r \sim \sqrt{R/h}, \\ 
\frac{ \kappa}{X^2} N^{4} \lambda^{-1} 
\tilde{\sigma}_{\delta} &
r \sim \sqrt{h/R} N^{2}, N \ll \sqrt{R/h}.
\end{cases}
\label{eq:expectedk}
\end{equation}
Clearly, for $\lambda \le 10^{-3}$ the factor of $\lambda^{-7/3}$
will become very large, so the shortest wavelength
buckling modes will approach zero very
quickly on the scale of the ridge parameters. Even the lower value of 
$\lambda^{-1} N^{4}$ should be a 
pronounced feature in the normal mode
spectrum for any $N$ of order unity. 


We can also calculate the possible
wavelengths of the buckling modes on the ridge-line
by substituting the scaling dependencies into
the two limits of $r$ presented in Eq.~\ref{eq:tworroots}. 
The high wavenumber cylinder buckling mode has
$r \sim \sqrt{R/h} \sim \lambda^{-1/3}$, while the low wavenumber mode 
has $r \sim \sqrt{h/R} N^{2} \sim \lambda^{1/3} N^{2}$. Thus, the 
limiting wavelengths of the classical buckling mode are
\begin{equation}
\zeta_{cl} = \frac{2 \pi R}{r} \sim 
\begin{cases} 
2 \pi X \lambda^{2/3} &
r \sim \sqrt{R/h}, \\
2 \pi X N^{-2} &
r \sim \sqrt{h/R} N^{2}, N \ll \sqrt{R/h}.
\end{cases} 
\label{eq:bucklewavelength}
\end{equation}
So for an aspect ratio of 
$\lambda=10^{-3}$ the
shortest buckling wavelength is of order $X/20$. 
The longer wavelength is independent of $\lambda$. 
On the ridge, the transverse wavenumber $N$ does not have a lower bound, 
but the longitudinal wavelength $\zeta_{cl}$ will presumably be a
half-integer fraction of the ridge length. 
We therefore expect that the longest
of these buckling wavelengths will be on the
order of the ridge length.
For $\zeta_{cl} = X$, the 
corresponding value of $N$ is $N = \sqrt{2\pi} \approx 2.5$.

Numerically
we looked for the buckling mode among the lowest normal modes of our 
simulated sheets for equilibrium configurations from zero forcing up 
to the buckling threshold. We found the modes by analytically 
calculating the the matrix of second derivatives of the total 
elastic energy
for the equilibrium 
positions of the sheet, then using a 
block-Lanczos algorithm~\cite{matrixmanip} to find 
several of
the lowest eigenmodes of this matrix. The eigenvalue corresponding to 
the buckling mode is precisely the spring constant $K_{\delta}$ 
defined above.
We used the 
``Underwood'' implementation of the block-Lanczos algorithm,
which is freely available on the NetLib online 
archive~\cite{netlib}. The block-Lanczos method is efficient at 
finding extremal eigenvalues and eigenvectors for large sparse 
matrices -- our matrices were large by virtue of the large lattice 
size,
but sparse since local curvature and strain fields at any lattice 
point are determined by relative positions of other grid points only 
up to a distance of next-next-nearest neighbors. The numerical values of 
elements in our second derivative matrices for very thin sheets 
differed by up to four orders of magnitude, so convergence of the 
Underwood routine was slow, taking up to several days 
to recover eight eigenmodes on a 700MHz Linux-based computer.

\begin{figure}

\center

\epsfig{file=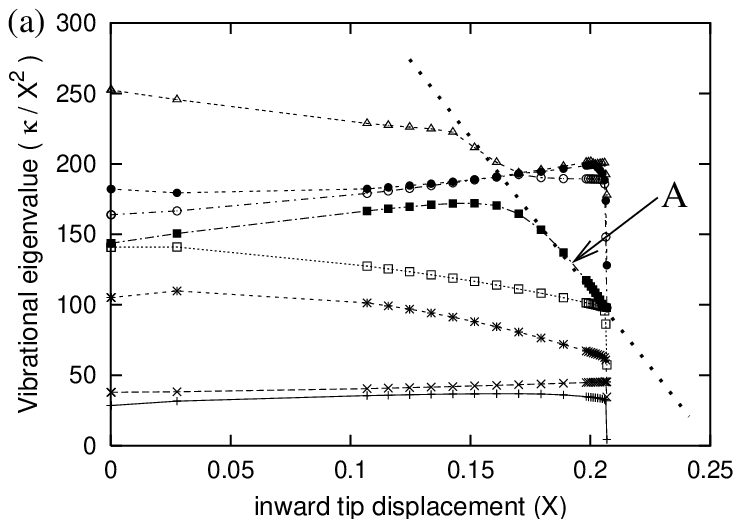}

\epsfig{file=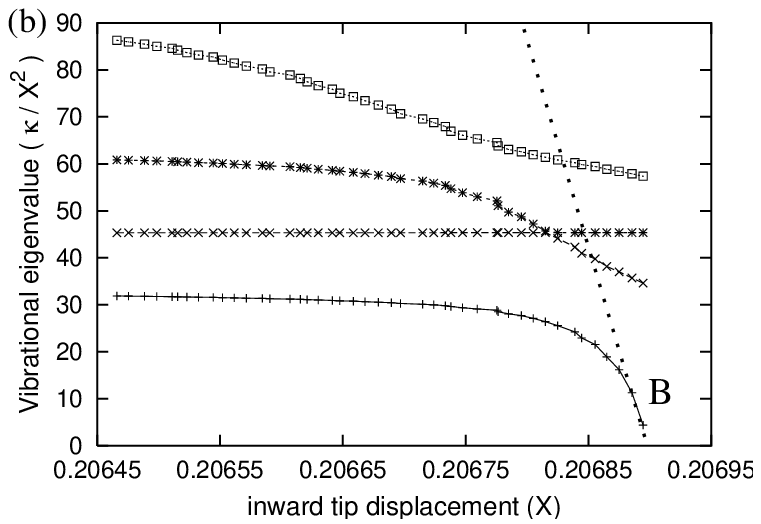}

\caption{Evolution of modes}
\label{fig:eigenvstt}

\legend{Both graphs plot values of the effective spring 
constants $K_{\delta}$
as a function of ridge tip displacement for
eigenmodes of a ridge with aspect ratio 
$\lambda=2 \times 10^{-3}$. The
ridge tip displacement is nearly linear in the ridge strain and stress,
so for these graphs $\tilde{\sigma}_\delta \approx (0.2 - \Delta)/0.2$.
The top graph plots the eight lowest eigenvalues 
at several different ridge tip displacements
(under application of inward external forces at the tips). The dashed 
line in (a) has a slope of approximately $4
\times 10^2 \frac{\kappa}{X^2 \tilde{\sigma}_\delta}$. The lower
graph is a closeup of the lowest four eigenvalues very near the buckling
threshold. The dashed 
line in (b) has a slope of approximately $10^6 
\frac{\kappa}{X^2 \tilde{\sigma}_\delta}$ (for comparison to scaling
values, $\left(0.002\right)^{-1} = 5 \times 10^2$ and
$\left(0.002\right)^{-7/3} = 2 \times 10^6$). The labeled points (A) and
(B) correspond to the modes pictured in Figure~\ref{fig:eigenmodes}(a)
and (b) respectively.}

\end{figure}

\begin{figure}

\center

\epsfig{file=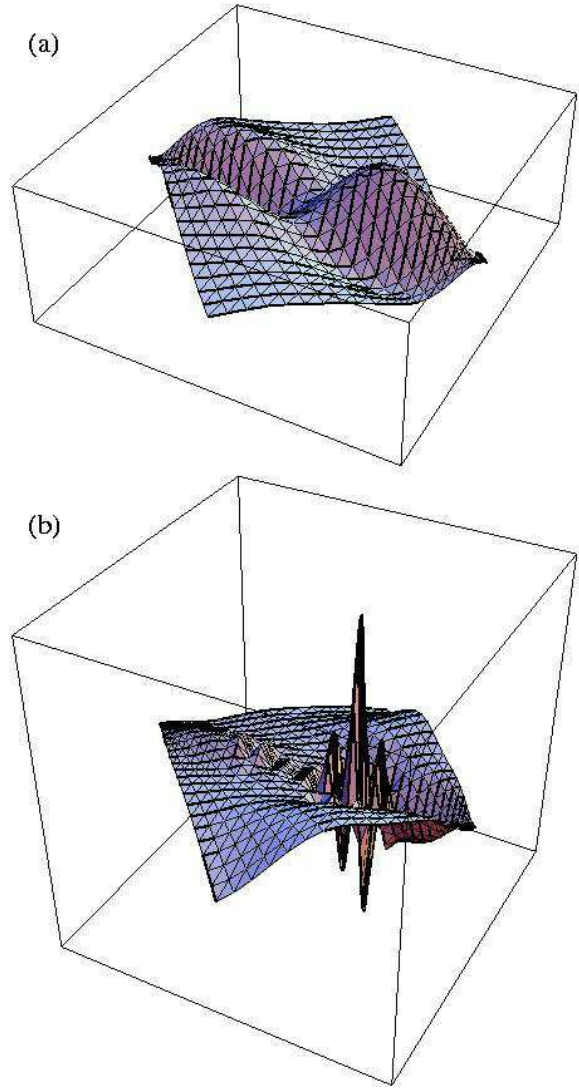, width=3in}

\caption{Representative eigenmodes}
\label{fig:eigenmodes}

\legend{In both images, the $x$ and $y$ coordinates are the material
coordinates of the sheet, while the $z$ coordinate is the eigenmode
motion normal to the ridge surface. In these sheets the ridge-line 
extends from the upper left to lower right corners.
The top image shows a longer wavelength mode 
which covers the length of the ridge. The the eigenvalue and tip
displacement for this mode are label by point (A) in
Figure~\ref{fig:eigenvstt}(a). 
The bottom image
shows a short wavelength mode. The the eigenvalue and tip
displacement for this mode are label by point (B) in
Figure~\ref{fig:eigenvstt}(b). }

\end{figure}

Figure~\ref{fig:eigenvstt} shows the evolution of the eight lowest 
eigenvalues as a function of vertex displacement for a ridge with 
aspect ratio $\lambda=2 \times 10^{-3}$. The eigenvalue evolution is 
qualitatively the same for thinner sheets as well. Over a
large range of inward vertex displacement, the
lowest modes all have nearly constant eigenvalues. The modes contain an
assortment of motions which are either global or localized on the
boundary layer or the ridge flanks. As the vertex displacement (and
therefore the ridge stress) is increased, the eigenvalues
corresponding to long wavelength
modes localized on the ridge begin to drop
more steeply. For example, the strongly sloped line in the upper 
right corner of Figure~\ref{fig:eigenvstt}(a)
corresponds to the mode shown in
Figure~\ref{fig:eigenmodes}(a). 
The eigenvalue for this mode
has a slope on the order of $-4 \times 10^2$ 
(in units of $\frac{ \kappa}{X^2} / \tilde{\sigma}_{\delta}$),
which is just one order greater than the minimum slope
predicted by
Eqs.~\ref{eq:expectedk}~and~\ref{eq:bucklewavelength}
for the
$\lambda^{-1} N^{4}$ scaling associated with long
modes of wavelength $X$ ($N = \sqrt{2 \pi}$). Similar modes, with
wavelengths $2X/3$, $X/2$, $2X/5$, etc., were found higher in the
eigenmode spectrum. Near the buckling threshold, these modes were also
seen to approach zero eigenvalue with slopes greater than, but on the
order of, that for the wavelength $X$ slope described above. The
computational time required to calculate higher modes prevented us from
studying them in greater detail, but they behaved
fundamentally the same as the mode shown in 
Figure~\ref{fig:eigenmodes}(a). 

None of these modes ever reaches zero value before the ridge
buckles, however -- instead a very localized, short wavelength
mode like that pictured in 
Figure~\ref{fig:eigenmodes}(b) appears suddenly, with very sharply
dropping eigenvalue, just before the ridge buckles. The evolution of 
the eigenvalue associated with this mode is shown in 
Figure~\ref{fig:eigenvstt}(b). The short wavelength mode
has a wavelength on the order of the lattice spacing and is asymmetric
about the center point of the ridge. For ridges 
with aspect ratio $\lambda= 2 \times 10^{-3}$, 
Eq.~\ref{eq:bucklewavelength} gives
a minimum wavelength of approximately $X/10$, which is on the 
same order as the
mid-ridge local lattice spacing. Thus the 
observed short wavelength mode is on 
the order of the minimum allowed
wavelength, and should therefore have a spring constant near that
predicted by Eq.~\ref{eq:expectedk} for $\lambda^{-7/3}$ 
scaling of $K_{\delta}$. As 
Figure~\ref{fig:eigenvstt}(b)
shows, the final slope of the eigenvalue as it approaches zero is indeed
on the right order of magnitude to fit the cylinder buckling theory.

%

\begin{figure}

\center 

\epsfig{file=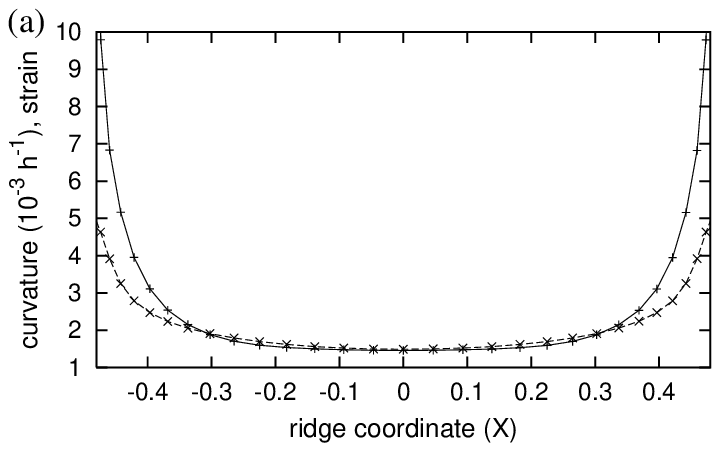}

\epsfig{file=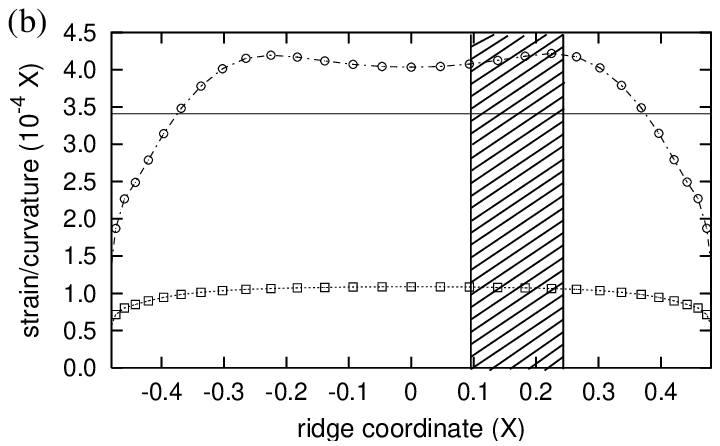}

\epsfig{file=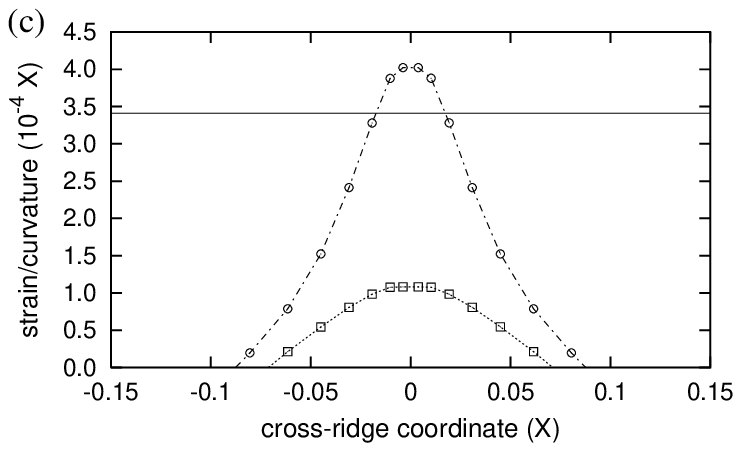}

\caption{Curvature and strain profiles on ridge}
\label{fig:nonuniform}

\legend{Plot (a) shows the curvature in units of $h^{-1}$ (``$+$'' symbol)
and strain magnitude (``$\times$'' symbol)
as a function of position along the ridge-line for a
ridge at its buckling threshold. 
The ridge has a thickness aspect 
ratio of $\lambda=5.5 \times 10^{-4}$. Plot (b) shows the ratio of
curvature to strain magnitude along the ridge-line for the same ridge at
rest (``\mybox'' symbol) and at the buckling 
threshold (``$\circ$'' symbol). The location and extent
of the localized
vibrational mode from Figure~\ref{fig:eigenmodes}(b) 
is highlighted.
Plot (c) shows the ratio of
curvature to strain magnitude across the middle of the ridge 
for the same ridge at
rest (``\mybox'' symbol) and at the buckling 
threshold (``$\circ$'' symbol). The horizontal lines in (b) and (c) 
are the
classical buckling value of the strain to curvature ratio predicted by
Eq~\ref{eq:cylbreakratio}. }

\end{figure}

Both the long and short wavelength
modes seem to obey the scaling
of cylinder buckling modes as they approach instability. 
That the short wavelength mode
reaches zero eigenvalue first in every case could be due to some
suppression of the long mode, 
either by boundary conditions or by the
changing geometry of the boundary layer along its length. 
Also, the short wavelength mode may well be 
enhanced by lattice effects, and
therefore should be more prone to cause buckling. 
The short wavelength mode is also enhanced by its high localization -- 
since the stress and curvature 
are not uniform along the ridge (see Figure~\ref{fig:nonuniform}), 
localized
patches of the ridge-line will meet the stress to curvature
threshold criterion, Eq.~\ref{eq:cylbreakstress}, 
before it is satisfied globally. 
In any case, the theory
developed for cylinders at the beginning of Section~\ref{ch:buckling}
allows for both
these families of modes approaching zero spring constant
just before the ridge buckles. The observations of these modes and their
matching eigenvalue slopes strengthens the connection between ridge and
cylinder buckling. 

The spatial extent along the ridge of the
short wavelength mode envelope
was observed to be independent of the sheet thickness.
The longitudinal 
wavelength of the buckling was at the lattice spacing. 
Figure~\ref{fig:nonuniform}(b) shows 
that the strain to curvature ratio on the mid-line itself surpasses 
the classical buckling threshold value by nearly $20 \% $. 
However, as Figure~\ref{fig:nonuniform}(c) shows,
this ratio drops away quickly in the direction
transverse to the ridge-line. Thus the line
plotted in Figure~\ref{fig:nonuniform}(b) is a very localized 
maximum profile. 
The buckling envelope has a width on the order of the ridge width, and 
the strain to curvature ratio
should be super-critical over this width 
before buckling occurs.
The minimum cylinder buckling 
wavelength shrinks 
as $\lambda^{2/3}$ as the sheet gets thinner, while the ridge width $R$
shrinks as $\lambda^{1/3}$, so the cylinder buckling mode should be
increasingly dependent only on local curvature and strain fields as 
$\lambda \rightarrow 0$. Therefore the short wavelength
localized mode should continue to
be the preferred buckling mode for thinner sheets. Numerically the 
time required to compute eigenmodes grew very quickly as we decreased 
the thickness aspect ratio $\lambda$, so we were not able to track the 
evolution of the short wavelength modes for significantly thinner 
sheets.


\subsection{Other geometries}

\subsubsection{Different ridge angles}
\label{sec:diffangles}

\begin{figure}

\center

\epsfig{file=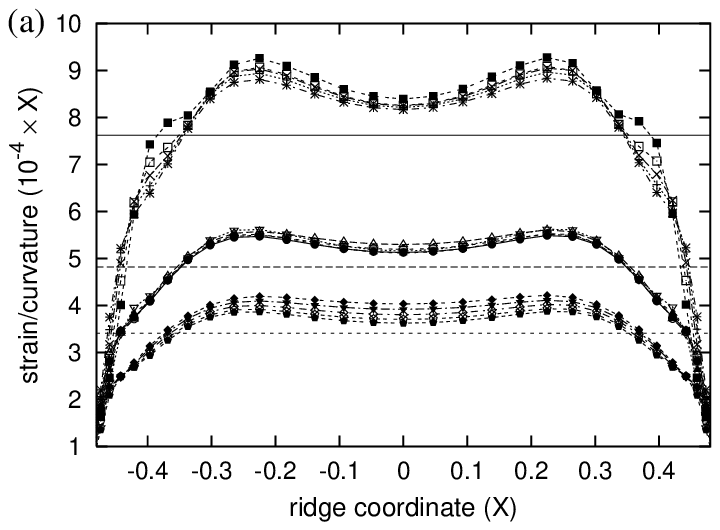}

\epsfig{file=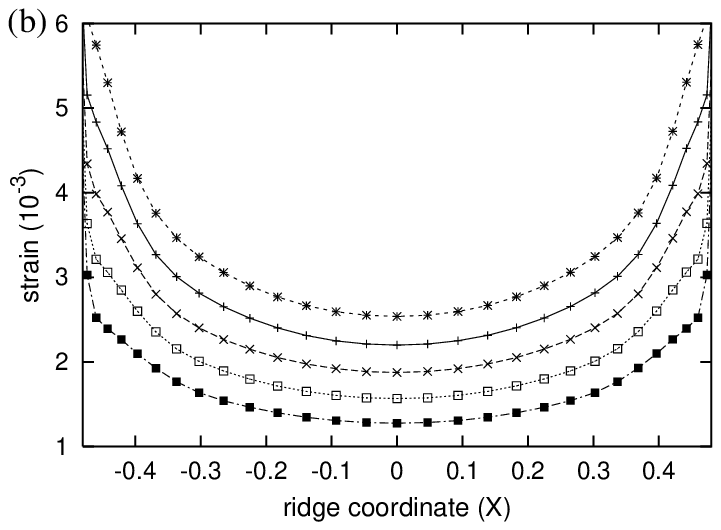}

\caption{Curvature and strain profiles}
\label{fig:otherangles}

\legend{Plot (a) shows the ratio of strain magnitude to curvature
along the ridge-line at the buckling threshold for five different 
dihedral angles and three different thicknesses.
The ridges had dihedral angles ranging from $\frac{\pi}{2}$ to 
$\frac{7\pi}{10}$. Regardless of ridge angle, the values of the
strain/curvature ratio tend to be grouped for ridges with the same
thickness aspect ratio. In the top grouping, 
$\lambda=1.25 \times 10^{-3}$, in the
middle $\lambda=8 \times 10^{-4}$, and for the 
bottom grouping $\lambda=5 \times 10^{-4}$.
Plot (b) shows the buckling threshold values of the strain magnitude
along the ridge-line for ridges 
with $\lambda=1.25 \times 10^{-3}$ and 
dihedral angles ranging from $\frac{\pi}{2}$ (top line) to 
$\frac{3\pi}{10}$ (bottom line). The horizontal lines in (b) are the
classical buckling values of the strain to curvature ratio predicted by
Eq~\ref{eq:cylbreakratio} for each thickness.}

\end{figure}

Most of our in depth buckling mode data analysis has been performed 
on ridges with dihedral angle $\frac{\pi}{2}$. 
However, the equation 
we derived for the critical stress (Eq.~\ref{eq:cylbreakstress}) 
only depends on two parameters of the ridge shape -- 
the transverse curvature and the material thickness. In order to show
that this relation holds for more general 
geometries, we again consider ridges with different dihedral 
angles. As in Section~\ref{app:simscaling}, our simulations used
sheets with the same size and length to width ratio, but the location
and orientation of the reflective planes for the edge boundary
conditions were changed (the connection between reflective planes and
dihedral angle is illustrated in Figure~\ref{fig:simfig}). 
The energy of
ridges with dihedral angle $\pi - 2\alpha$ scales 
as $\alpha^{7/3}$, so adjusting
this angle changes the curvature across the ridge significantly. Still,
as Figure~\ref{fig:otherangles} shows
for ridges with dihedral angles ranging
from $\frac{\pi}{2}$ to $\frac{7\pi}{10}$ that the ratio of 
strain to curvature along the
ridge at the buckling threshold was the same for ridges with the same
aspect ratio $\lambda$. This is true despite the variation in the
buckling strain by a factor of $2$ between the largest and smallest
angled ridges. This further affirms the cylinder buckling hypothesis.

This observation may also explain the insensitivity of the observed
fractional change in energy between resting ridges and those at the
buckling threshold. Another consequence of Lobkovsky's treatment 
in~\cite{Alex} is that the longitudinal 
strain and transverse curvature on the ridge-line scale with the same 
power of $\alpha$. Since our ridges always 
buckle at the same value of the ratio of strain to curvature, the 
identical scaling of these quantities with $\alpha$ implies an 
identical fractional change in their values (and therefore the total 
energy of the ridge) between the resting and 
buckling threshold states.
We expect this behavior to extend to dihedral angle up to $\pi$, though
we haven't demonstrated it.

\subsubsection{Longer ridge flanks}
\label{sec:longerflanks}

\begin{figure}

\center

\epsfig{file=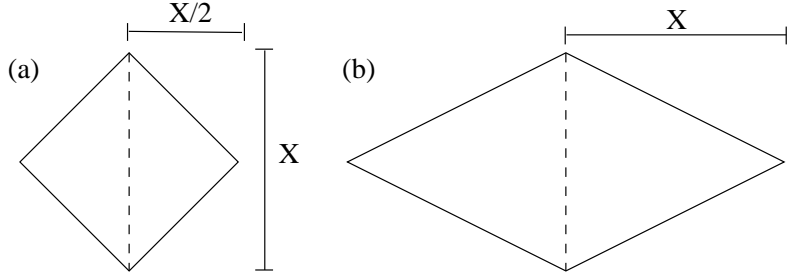}

\caption{Shape of simulational grids}
\label{fig:flankstretch}

\legend{The dashed lines indicate the location of the simulated
ridge-line. The grid in (b) was used for the simulations described in
Section~\ref{sec:longerflanks}. The grid in (a) was used for all other 
ridge simulations.}

\end{figure}

\begin{figure}

\center

\epsfig{file=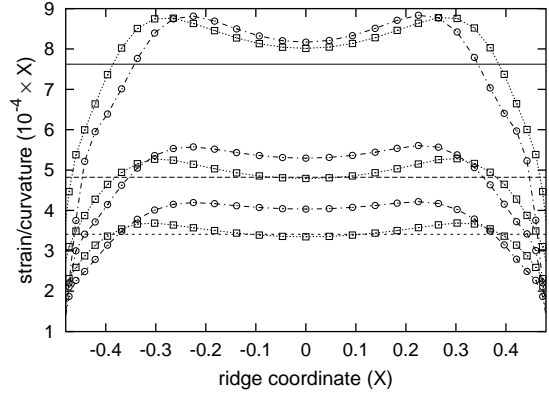}

\caption{Strain to curvature ratios for ridges with longer flanks}
\label{fig:flankratios}

\legend{The strain to curvature ratios on the ridge-line at the 
buckling threshold are shown for our typical ridges (``$\circ$''
symbols) and for ridges with longer flanks (``\mybox'' symbols).
Thickness aspect ratios $\lambda$ shown here range from
$1.25 \times 10^{-3}$ (top curves) to 
$5.6 \times 10^{-4}$ (bottom curves).
The horizontal lines are the
classical buckling values of the strain 
to curvature ratio predicted by
Eq~\ref{eq:cylbreakratio} for each thickness.}

\end{figure}

\begin{figure}

\center

\epsfig{file=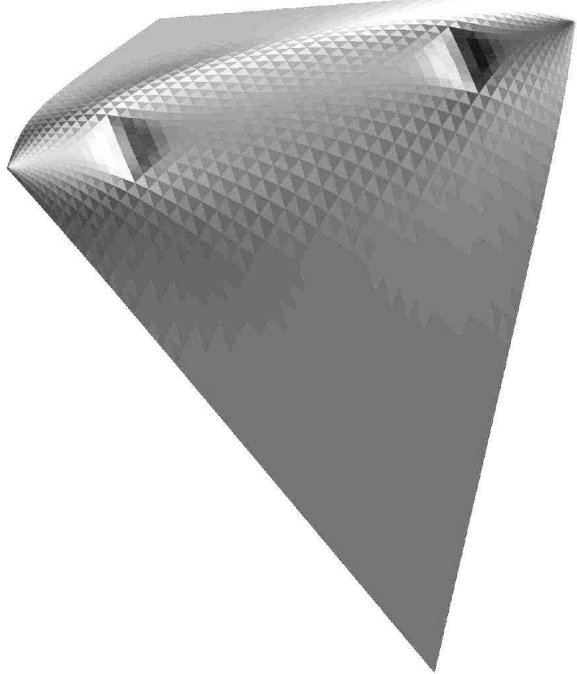, width=3in}

\caption{Buckling pattern on ridge with long flanks}
\label{fig:flankbuckle}

\legend{This ridge had a
thickness aspect ratios $\lambda$ of
$1.25 \times 10^{-3}$. Lighting and 
shading were chosen to emphasize physical features.}

\end{figure}

So far, we have simulated ridges on a grid whose edges formed a perfect
square, as shown in Figure~\ref{fig:flankstretch}(a). 
The ridge-line extended between two corners of this grid, so its
simulated flanks were right triangles.
The non-vertex corners of the grid were a distance 
$X/2$ from the center of the ridge-line, where $X$ is the ridge length. 
As a variation on this
ridge geometry, we also simulated ridges with flanks which were twice as
long as those for the typical simulations, using the grid shown in  
Figure~\ref{fig:flankstretch}(b). On this grid 
the non-vertex corners were
a distance $X$ from the center of the ridge-line. 
The boundary conditions were again reflective planes which were oriented
to give the ridge a $\frac{\pi}{2}$ dihedral angle.

We simulated ridges with thickness aspect ratio $\lambda$ ranging from
$1.25 \times 10^{-3}$ to $1.25 \times 10^{-4}$.
For these ridges we once again found that the buckling threshold energy
was approximately $\universal$ greater than the resting ridge energy.
As Figure~\ref{fig:flankratios} shows, we also found that the ratio of
strain to curvature along the ridge-line at the buckling threshold was
nearly equal to the threshold values for our typical ridges. 

Interestingly, the peaks in the strain-curvature ratio
along the ridge-line were consistently closer to the
vertices for this geometry than they were for all the other geometries
we studied (this is visible in the profiles
shown in Figure~\ref{fig:flankratios}).
This suggests that the
locations of these peaks are determined by boundary conditions. The
appearance of the peaks is probably an effect of the ridge pulling on
its mirror image. 
Predictably, the change in
location of this peak also causes the ridge to buckle closer to its
vertices, as shown in Figure~\ref{fig:flankbuckle}. This is strong
evidence that buckling occurs near the first
localized patch on the ridge-line where the strain-curvature ratio is
super-critical. 
When this pulling is absent, we might expect the
buckling region to move towards the center. Such a case occurs in
Lobkovsky et al.'s minimal ridge~\cite{our.stuff}, 
where there is no applied stress
at the boundaries, and buckling occurs at the center of the ridge.
The change in buckling location is also consistent 
with the observed locations of new ridges on a
buckled tetrahedron in~\cite{our.stuff}. 
Since the tetrahedron ridges
have shorter flanks than the cube we use, the new ridges may be expected
to form closer to the center of the ridge on the tetrahedron. 

\subsubsection{Shell buckling}

We have derived the behavior of the ridge buckling mode near
the buckling transition as a function of the {\it total} longitudinal
stress on the ridge-line, with as little reference as possible to where
this stress comes from. At this point, we wish to 
address the role of the resting ridge stress in the buckling
transition. As we showed in Section~\ref{ch:prebuckle}, even in its
``resting'' state the ridge has significant longitudinal stress -- the 
ridge
stores a nonzero fraction its total elastic energy in the stress field.
The amount of work required to buckle the ridge is therefore
only that required to 
increase the longitudinal stress from the resting value
to the buckling threshold value.
It seems an odd occurrence that the energy required to make
the ridge should also do part of the work required to break it. It
is even more intriguing since the scaling of ridge stress and critical
stress are identical -- if this were not so then any ridge above or
below (depending on the relative scaling) a critical length would buckle
spontaneously.\footnote{Spontaneous buckling is commonly
observed numerically due to
lattice effects in very thin sheets. Any departure from the scaling
laws due to lattice size in simulations
or irregularities in real materials
could be sufficient to push the required
resting ridge stress over the buckling threshold.} As it is, there are
most likely some geometrical constraints placed on allowed ridge
configurations purely by virtue of the fact that the ridge stress is
naturally on the same scale as the buckling stress.

To find out whether or not ridge stress weakens the ridge, we
numerically studied 
how the buckling threshold changed when the resting ridge stress was
removed. To do this, we first found the
minimum energy configuration of a resting ridge
for a given thickness aspect ratio $\lambda$. 
We then redefined all the
lengths and curvatures in the sheet such that the resting ridge
configuration had zero stain and curvature, and thus
zero resting energy.
If we denote the strains and curvatures of the resting ridge as
$\gamma^o_{ij}$ and $C^o_{ij}$ respectively, we can write 
distortions away from this state as
\begin{gather}
\gamma'_{ij} = \gamma_{ij} - \gamma^o_{ij} \\
C'_{ij} = C_{ij} - C^o_{ij}, 
\end{gather}
where $\gamma_{ij}$ and $C_{ij}$ are computed as before.
These primed quantities were substituted into 
the strain and curvature energy equations, 
Eqs.~\ref{eq:stretchen} and~\ref{eq:benden}, to make the 
energy for the resting ridge
configuration identically zero. We refer to 
sheets which have
intrinsic curvature and non-flat metrics as a shells.

\begin{figure}

\center

\epsfig{file=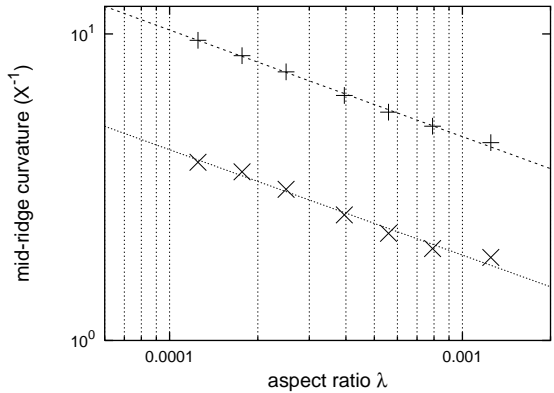}

\caption{Shell vs. ridge buckling energies}
\label{fig:shellbuckleenergy}

\legend{This graph shows the difference between resting and buckling
energies for shells (``$+$'' symbols) and ridges (``$\times$'' symbols).
The numerical scaling fits had exponents consistent
with $-1/3$.}

\end{figure}

\begin{figure}

\center



\epsfig{file=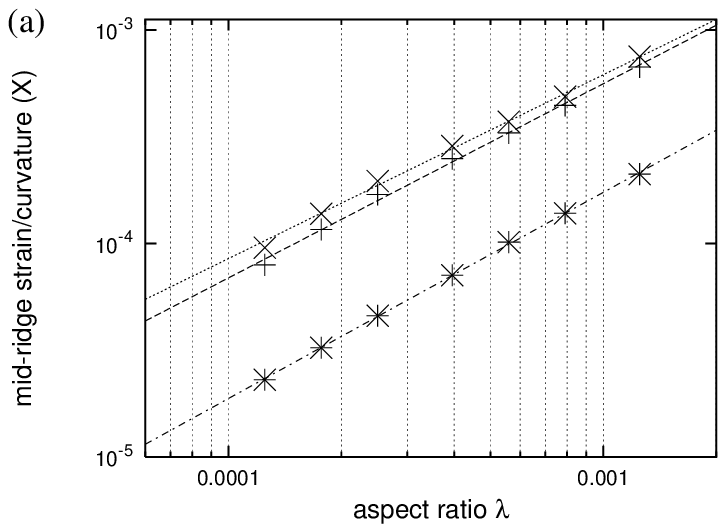, width=3in}

\epsfig{file=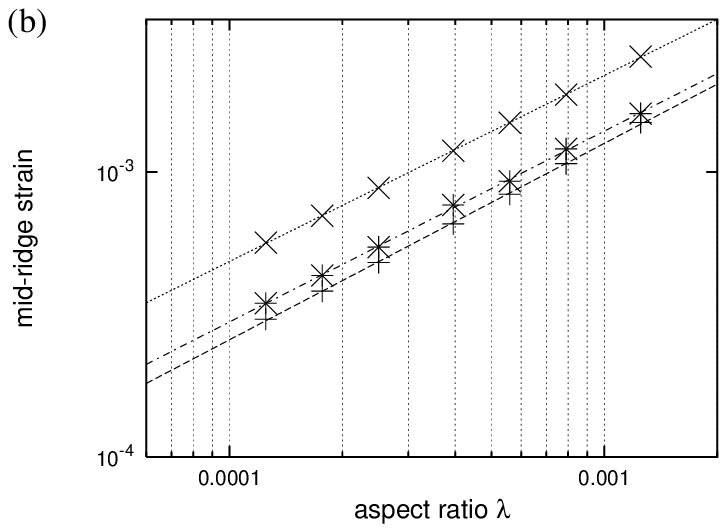, width=3in}

\caption{Shell vs. ridge buckling configurations}
\label{fig:shellbucklegraph}

\legend{
In each graph, 
``\crossstar'' symbols denote values for ridges at rest,
``$\times$'' symbols denote values for ridges at their buckling
threshold, and
``$+$'' symbols denote values for shells at their buckling
threshold. The numerical scaling fits in (a) had exponents consistent
with $1$ while those in (b) had exponents consistent
with $2/3$.}

\end{figure}

\begin{figure}

\center

\epsfig{file=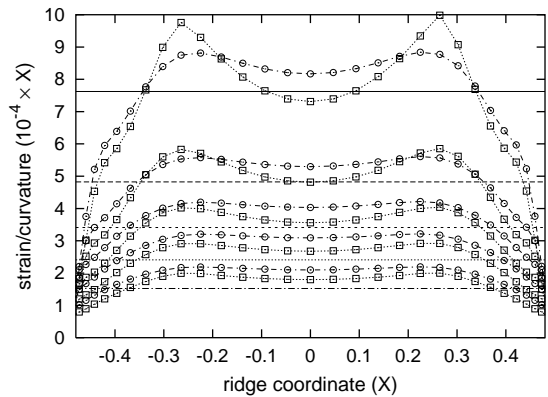}

\caption{Strain to curvature ratios for shells}
\label{fig:shellratios}

\legend{The strain to curvature ratios on the ridge-line at the 
buckling threshold are shown for ridges (``$\circ$''
symbols) and shells (``\mybox'' symbols) with thickness ratios
$\lambda$ ranging from $1.25 \times 10^{-3}$ (top curves) to 
$2.5 \times 10^{-4}$ (bottom curves).The horizontal lines are the
classical buckling values of the 
strain to curvature ratio predicted by
Eq~\ref{eq:cylbreakratio} for each thickness.}

\end{figure}

As with the ridges before, we buckled the shells by imposing a gradually
increasing hard wall potential at the vertices. 
We found that for any given thickness,
it takes more work to buckle the shell than it does the corresponding
ridge, though each started with exactly the same geometry 
(see Figure~\ref{fig:shellbuckleenergy}). 
Though the two systems evolve differently from their
initial states under the applied load, 
Figure~\ref{fig:shellbucklegraph}(a) confirms that
each buckles at nearly the same ratio of {\it
total} stress to {\it
total} (intrinsic plus extrinsic) curvature. 
This is just the result
that would be predicted from the cylinder theory in
Sec.~\ref{sec:cylinders}, since that buckling mode depends only on the
total value of the radius of curvature $R$ as a geometric quantity, 
without reference to its energetic cost. 
Figure~\ref{fig:shellratios}
shows that the strain to curvature ratios
along the ridge-line are nearly the same for ridges and shells with
identical thickness aspect ratios $\lambda$. The figure also shows that
the maximum of this ratios is near the same place on both ridges and
shells, though it is more sharply peaked on shells.

Also, Figure~\ref{fig:shellbucklegraph}(b) shows that although the
shell cannot sustain the total amount of longitudinal stress that the
ridge holds at the buckling threshold, it can sustain more 
{\it additional} stress, starting from the resting ridge geometry, than
the ridge before it buckles. The reason that the shell cannot sustain
the same total stress is that it grows much flatter with applied load
than does the ridge, so it reaches the critical stress to curvature
ratio at a lower value of both these quantities. The difference in the
evolution of the curvature with applied loading is due just to the
fact that the cross-ridge curvature is determined by an energy balance,
and the energetic term are much different for these two systems.

\subsection{Universality of buckling energy}
\label{sec:universality}

It is striking that for all the ridge geometries we studied,
the buckling
threshold energy was approximately $\universal$ 
greater than the resting
energy. Our present ridge theory is not sufficient to
fully explain this ratio, but we take this opportunity
to speculate about how universal it may be.

In~\cite{our.stuff}, Lobkovsky et al. 
stated that the bending and stretching
energies on ridges should obey a virial relation, with total bending
energy five times greater than total stretching. Assuming that both
energies are only significant on the ridge-line, then the virial
relation also extends to typical bending and stretching energy
densities. Taking the largest terms from
Eqs.~\ref{eq:stretchen}~and~\ref{eq:benden}, this gives:
\begin{equation}
\frac{Yh^3}{12 \left( 1 - \nu^2 \right)} C_{yy}^2 \approx
\frac{5 Yh}{\left( 1 - \nu^2 \right)} \gamma_{xx}^2,
\end{equation}
which reduces to
\begin{equation}
\gamma_{xx} \approx \frac{1}{\sqrt{60}} h C_{yy} 
\approx 0.13 \ h C_{yy}.
\end{equation}
From Eq~\ref{eq:cylbreakratio}, the classical value of the 
breaking strain for $\nu =1/3$ is
\begin{equation}
\gamma_{cl} = \approx 0.61\ h C_{yy}.
\end{equation}
In our simulations we observed that the bending energy doesn't change
significantly as the ridge-line is compressed by the applied force. 
If the ridge curvature stayed the same and the ridge strain increased
from its resting value up to the classic value, then the ratio of
resting energy to buckling threshold energy would be
($5+1$):($5+\left(0.61/0.13\right)^2$).  The breaking energy would be
approximately $4.6$ times the resting energy.  However,
Figure~\ref{fig:rescaling} shows that the mid-ridge curvature
decreases as force is applied.  Between resting and buckling, the
curvature decreases by nearly a factor of two.  The effect of this
flattening of the ridge-line is to decrease the breaking strain by a
factor of two and the breaking stain energy by a factor of four.  The
corresponding breaking energy is only $1.77$ times greater than the
resting energy.

The discrepancy between the predicted factor $1.77$ and the observed
factor of $\oneplusuniversal$ is understandable, 
given the simplicity of
our approximations. In reality the distribution of strain and
curvature on the ridge-line are not identical.  In real ridges we may
expect the local ratios to vary by factors of order unity depending on
boundary conditions.  However, we presently cannot 
explain why the local
curvature at the center of the ridge-line drops by a factor of two
while the total bending energy remains constant. This factor of two 
seems to be universal throughout our simulations, and we predict 
that the curvature won't drop by significantly larger fractions for 
different ridge boundary conditions. Still, a detailed understanding
of this factor remains an open question.

\subsection{Discontinuity at buckling}

\begin{figure}

\center

\epsfig{file=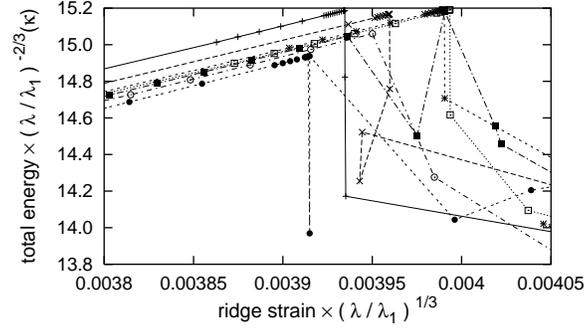}

\caption{Post-buckled energy}
\label{fig:postbuckleen}

\legend{This plot shows 
the total rescaled elastic energy of several simulated
ridges as a function of rescaled inward vertex displacement just before
and after the ridges buckled. The ridges
had aspect ratios $\lambda$ ranging
from $5 \times 10^{-4}$ to $5 \times 10^{-5}$. 
According to the scaling analysis in Section~\ref{ch:prebuckle}, 
the rescalings of $(\lambda/\lambda_1)^{-2/3}$ for energy and
$(\lambda/\lambda_1)^{1/3}$ for vertex displacement would collapse all
the lines on to one for perfect ridge scaling. The observed
discrepancies between the pre-buckling part of the
lines is small on the scale of the entire ridge
evolution. In these graphs $\lambda_1 = 5 \times 10^{-4}$.}

\end{figure}

To complete our study of the buckling transition, we comment briefly
on the post-buckled state and its rapid growth from the unbuckled
state.  As mentioned above, the equilibrium configuration immediately
after buckling contains at least one large additional ridge.  The
additional ridges appear suddenly -- when we first see them they are
already as long as the unbroken ridge was wide.  A significant change in
the elastic energy accompanies the transition, as shown in
Figure~\ref{fig:postbuckleen}.  It is also notable that the
post-buckled state bears little resemblance to the short wavelength
buckling mode which we credit with causing the transition.  Since the
\vonkarman equations are highly non-linear, the growth of the buckled
state quickly passes beyond the regime where it is well modeled by our
linear stability analysis.  We therefore presume that when the the
non-linear terms start to become important, they favor further growth
of the buckled state.  The net result is an energetic avalanche into a
completely different state.


\begin{figure}

\center

\epsfig{file=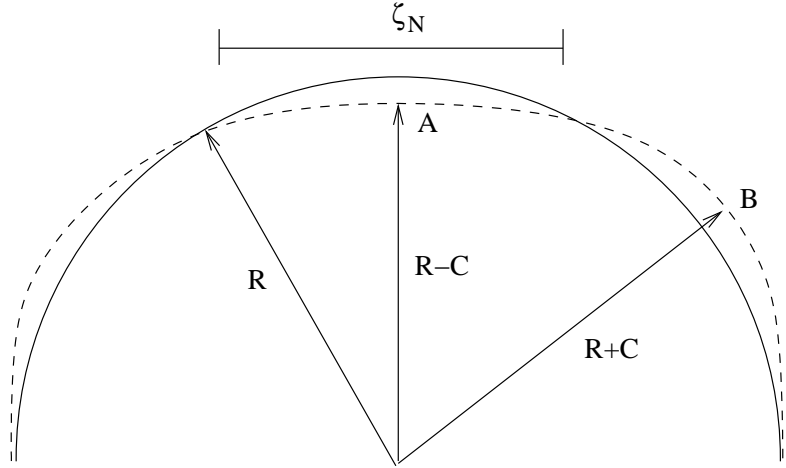}

\caption{Effect of buckling mode on $R$}
\label{fig:nonlinear}

\legend{The solid line represents an unbuckled
local patch of surface with
dominant radius of curvature $R$. The dashed line shows the 
post-buckled surface
with buckling mode amplitude $C$ and wavelength $\zeta_N$. From
Eq.~\ref{eq:curveonr}, the curvature at point A is approximately
$\frac{1}{R} - \frac{C}{\zeta_N^2}$ while the curvature at B is 
approximately $\frac{1}{R} + \frac{C}{\zeta_N^2}$. 
}

\end{figure}

\begin{figure}

\center

\epsfig{file=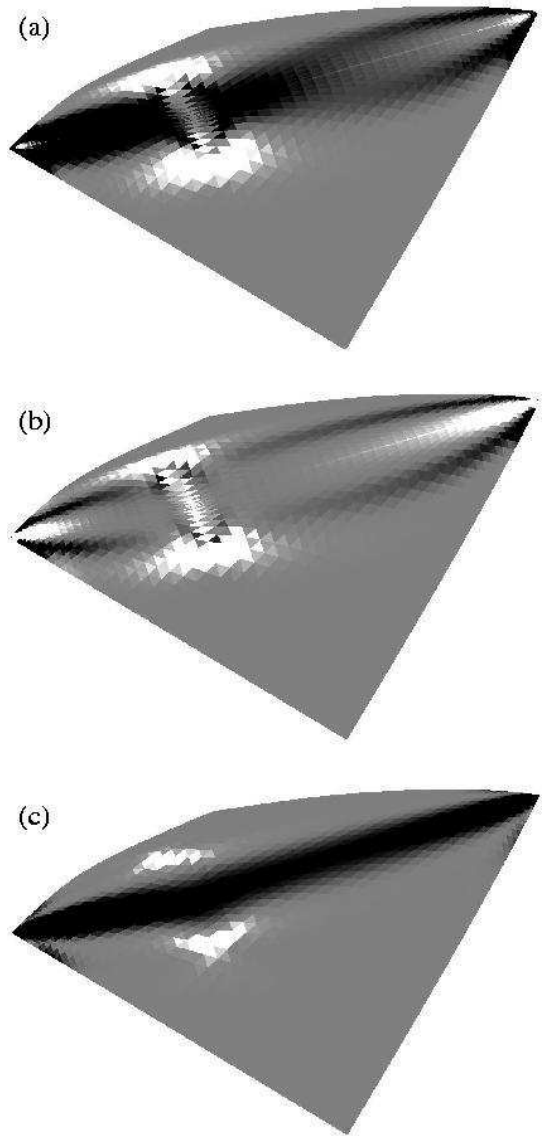, width=2.8in}

\caption{Redistribution of elastic energy at buckling}
\label{fig:redistribute}

\legend{Each plot shows the postbuckling configuration first shown in
Figure~\ref{fig:buckled}, with one large new ridge crossing the original
ridge. The original ridge had aspect ratio $\lambda=10^{-3}$. The plots
are shaded according to change in local elastic energy density between
the buckling threshold values and the postbuckled value for the pictured
configuration -- no
change is gray, increases are white, and decreases are black. Image (a)
shows the change in total elastic energy, image (b) shows the change in
bending energy, and image (c) shows the change in stretching energy.}

\end{figure}

As the buckling mode grows, 
it will begin to significantly perturb the
pre-existing stress and curvature fields $\sigma$ and $1/R$. 
Figure~\ref{fig:nonlinear} illustrates how the growth of the buckling
mode perturbs the large transverse curvature. 
From Eq.~\ref{eq:curveonr}, the maxima for which the displacement $w$
grows radially inward will decrease the local transverse radius 
of curvature by $-C/\zeta_N^2$, where $C$ is the buckling mode amplitude
and $\zeta_N$ is its transverse wavelength. The
maxima which grow radially outward will increase the local curvature by
the same amount.
We can also calculate the 
additional longitudinal stress due to the buckling mode itself
from Eqs.~\ref{eq:bucklemode}~and~\ref{eq:ABC} combined with
Eq.~\ref{eq:newstraindispl}:
\begin{eqnarray}
\sigma'_{xx} & = & \frac{Yh}{1-\nu^2} \left( \gamma'_{xx} + 
\nu \gamma'_{yy} \right) \nonumber \\
 & = & - \frac{Yh}{1-\nu^2} \left( \left( 1-\nu \right) 
+ \frac{\left(1 -\nu^2 \right)
r^2 N^2}{\left(r^2+N^2\right)^2} \right) \frac{C}{R} e^{irx/R}
\cos \left(Ny/R \right) \nonumber \\
 & \equiv &  - C \Theta \ e^{irx/R}
\cos \left(Ny/R \right) 
\label{eq:moresigma}
\end{eqnarray}
The coefficient $\Theta$ is always positive. 
Our frame is defined with positive normal displacements 
pointing inwards (downward and into the page 
in Figure~\ref{fig:redistribute}),
so Eq.~\ref{eq:moresigma} implies
that there is additional compression 
at the points of maximum inward deflection of the buckling pattern,
and matching extension at the
points of maximum outward deflection
(negative stress results from compression).
Thus, as the buckling mode grows, the local
ratio of strain to curvature becomes
\begin{equation}
\sigma/(\frac{1}{R}) \rightarrow
\begin{cases}
(\sigma + C \Theta )/(\frac{1}{R} - C/\zeta_N^2)
& \textrm{inward maximum}, \\
(\sigma - C \Theta )/(\frac{1}{R} + C/\zeta_N^2)
& \textrm{outward maximum}.
\end{cases}
\end{equation}
By the stability condition given in Eq.~\ref{eq:cylbreakratio}, the
inward growing maxima become more unstable to further growth while the
outward ones become less unstable.


From these simple arguments it is clear that the out-of-plane 
force balance changes dramatically as the buckling mode grows and 
non-linear terms become significant. Also, this reasoning indicates 
that the inward growing maxima are more favorable than outward growing 
ones. On a cylinder, the constraints of periodicity require an equal 
azimuthal number of
inward and outward maxima, so force balance is again achieved with the 
same number of maxima as the initial unstable mode. This constraint 
does not hold on the ridge, so inward maxima are free to grow into the 
region of the ridge flanks. The net result could be that 
one inward peak grows until it subsumes all the other maxima and 
becomes the one prominent feature of the buckled state -- a single 
large transverse ridge. 

The only reason for the 
growth of the new ridge to cease is that 
the potential energy driving this motion 
is exhausted. Figure~\ref{fig:redistribute} shows how the elastic 
energy redistributes itself upon the buckling of a ridge with aspect 
ratio $10^{-3}$. Predictably, the largest local
decrease in elastic energy is the loss of stress 
energy density along the ridge. In our buckling scheme 
it is the stored longitudinal stress which drives the growth of 
the buckling mode -- we observed that the inward buckling peak
which becomes the observed
single new ridge grows until the ridge-line stress is
nearly gone. Some of the energy is stored in the new ridge and new
vertices, but less than was stored in the threshold state.

Besides showing the jump in energy at buckling,
Figure~\ref{fig:postbuckleen} also shows the randomness in the observed
energies of the postbuckled state. Our simulations were optimized for
the pre-buckled ridges. Their accuracy in modeling the post-buckled
state is qualitative at best. There is no indication from this graph
that the post-buckled state has the same energy scaling as the
ridge. This is not surprising, since the ``resting'' configuration of
our new post-buckled state is not defined, 
so we cannot be sure that the states we see
immediately after buckling are at ``equivalent'' values of the tip
displacement, in the sense developed in Section~\ref{ch:prebuckle}
for a similarity solution at different material thicknesses.

Some excellent experimental work on the preferred crease size for a
circular cylindrical cross section under axial compression is presented
in~\cite{chieb.creases,chaieb.creases.exp}. In this work the authors found
that the actual saturation length of the new crease was determined by a
balance between the energy of the crease and that of the additional
singularities at its ends. 
In our geometry, the scaling of the new ridge energy
changes once it has grown beyond the width of the original boundary
layer. Therefore, the energy balance determined for the constant
curvature cross section may not be applicable. We leave this topic for
future research.


\section{Discussion}
\label{ch:discussion}

We have explored the behavior of a stretching ridge under the application 
of an external force potential. For the sake of clarity we have focused 
our simulations on a particular representative ridge geometry, but 
the response of this ridge has been shown to obey very general 
principles. In this section we re-cap our discoveries, putting them into
the broader context of the enhanced strength these spontaneous structures
add to thin sheets.
We also discuss the range of applicability for 
our approach to other ridge geometries, and to the behavior of 
collections of ridges in a crumpled sheet. Finally, we suggest engineering 
applications of the understanding we have gained concerning stretching 
ridges.

The resistance of materials to typical forms of distortion and damage
is a very well established field, with a history that dates back to the
19th century. However, a crumpled sheet derives its strength not just
from its material properties, but also from the spontaneous ridge
network it contains. This spontaneous network confers strength in a way
that clearly arises from the co-operative interaction between curvature
and strain. The novel aspect of this interaction has already been shown
by the newly identified scaling of the energy of these structures with
overall size of the system~\cite{our.stuff}. However, the strength
against collapse resulting from these structures has up to now been
poorly understood.
In a highly crumpled sheet, resistance to further deformation results
almost entirely from the work required to deform and break the ridges
which span the volume occupied by the sheet. The strength of ridges
in turn results from their shape, and their effective elastic modulus is
not related to the modulus of the component material in any simple
way. 

In Sections~\ref{ch:prebuckle}~and~\ref{ch:sims} we established a 
scaling relation for the response of a ridge to forces applied at its
endpoints. This is the type of forcing against which
ridges are strongest (have the highest effective modulus).
Presumably, when a force is applied in an arbitrary direction to a
moderately crumpled sheet, ridges which are oriented at broad angles to
the applied force will yield very quickly to it, and resistance to
the force will come from ridges which happen to be aligned parallel to
the forcing. Thus the ridge response to this particular 
forcing determines the effective elastic modulus
of crumpled sheets. We showed that,
given a knowledge of how one ridge of any size will respond to the
force at its ends, we can rescale the force to displacement relation to
all other ridge lengths by multiplying it by a simple power of the
thickness aspect ratio $\lambda$, namely $\lambda^1$. The force to
displacement relation for an individual ridge can be obtained through
simulations or simple estimates. Together with a model of the
distribution of ridge sizes in a typical crumpled sheet, our scaling
relation for the ridge strength gives a complete model for 
the effective elastic modulus
of the entire crumpled sheet, as well as the change in the sheet's
strength as it is further crumpled and the typical ridge sizes change.

We derived the scaling of the ridge force response by first
assuming that ridge scaling was still valid for forced ridges, 
then calculating the 
required rescaling of the perturbing force. Our approach is limited by 
the requirement that both the location and magnitude of applied 
forcing must be rescaled for a similarity solution. Also, there is no 
systematic way to determine if the scaling assumption will hold for 
every forcing. Still, our approach is comparable to other
treatments in its effectiveness, since
much of the physics of crumpling relies on 
intuition for each special case.

Happily, our approach was shown to be well suited for 
point forces applied to the vertices. For 
this case, the location of the forcing is fixed under 
rescaling, since the vertices by definition do not move when the 
ridge gets thinner. Also, since the applied forcing works almost 
entirely to compress the ridge-line, its coupling to the longitudinal 
ridge stress will be very strong. Thus the predicted 
scaling of the force response is unambiguous.

Forcing applied to other points on the boundary of the sheet will 
most likely not scale as cleanly as forcing
applied to the vertex. For other locations and angles, the 
applied force may result in large stress transverse to the ridge - 
it would therefore strongly perturb both transverse and longitudinal 
stresses. However, these stresses have different scalings on the resting 
ridge, so an equal perturbation of each would most likely ruin the 
ridge scaling. 

The other important perturbation, which we did not simulate,
is forcing applied normal to the 
surface. Scaling of this force response, as derived in 
Section~\ref{ch:prebuckle}, should be fairly robust since the term $P$ 
is perturbing a quantity which is zero for resting ridges. Therefore 
there is no pre-established scaling to destroy.

Our other important result, established in Section~\ref{ch:buckling},
was to link the buckling transition for 
ridges to the buckling of thin cylinders. This result
is supported by a great deal of analysis 
and is very general. Prior speculation held that ridges may derive 
anomalously large breaking strength from their pre-existing 
longitudinal strain.
We have shown that, in terms of strength, the ridge acts essentially 
as a cylinder whose radius scales with thickness. 
Whereas the work in Sections~\ref{ch:prebuckle}~and~\ref{ch:sims}
allowed us to understand the strength of crumpled sheets against small
deformations which did not change the structure of the crumpling
network, knowledge of the buckling strength lets us 
model the evolution of the strength and energy of a sheet throughout
the crumpling process, from the flat to the highly crumpled
state.

In Section~\ref{ch:buckling}, we show that the stability of the ridge
against buckling is determined completely by the local ratio of the
transverse strain to the longitudinal curvature on the ridge-line. Since
the transverse curvature on the ridge scales with its length, we can
immediately determine the buckling stress of any ridge as a function only
of its length and thickness. We
established that the allowed buckling wavelengths $\zeta_{cl}$
are between $2 \pi X \lambda^{2/3} \le \zeta_{cl} \le X$, where $X$ is
the ridge length and $\lambda$ is the thickness aspect ratio. Buckling
can take place when the strain to curvature ratio is supercritical over
a region larger than the minimum wavelength. We showed that ridges
buckle near the point at which this ratio has a 
localized maximum on the ridge-line. We established that the location of
this strain to curvature maximum does not depend on the dihedral
angle of the
ridge, but it does depend on the angle of the ridge-line relative to its
neighboring ridges. The simplicity of the buckling criterion established
here, as well as the clear connection between this ratio and the buckling
behavior of ridges, is a great improvement over the previous
understanding of the breaking strength of these structures. Further
development of the relation between a ridge and its neighbors may lead
to general laws regarding preferred distributions of angles separating
ridges in crumpled sheets -- this plus the length and energy
distributions discussed below could lead to an accurate 
statistical mechanics for crumpled sheets

It is our hope that the knowledge gained in this study can be
helpful in the development of a statistical mechanics for ridge
distributions in crumpled sheets. We have demonstrated for a range
of ridge angles under a typical form of forcing
that the energy at the buckling threshold is a fixed multiple of the
resting ridge energy. For our measurements this multiple was
approximately $\oneplusuniversal$ -- we 
speculate based on the arguments in Section~\ref{sec:universality}
that this multiple will not be greater
than $2$ in general. Combined with previous work which uncovered the
length and angle scaling of the ridge energy, this limitation on the
possible loading of a ridge should place some limitation on the relative
sizes of adjacent ridges in a typical crumpled sheet. This proposed
limitation follows from the assumption that in the interior of the
sheet, much of the force on vertices is carried through adjacent
ridges. Knowledge of the strength of ridges will definitely lend insight
to the evolution of successive crumpled states as a sheet is compressed.

Finally, in terms of applicability to real-world problems, 
the understanding of ridge buckling developed here
has practical import for the possible use of single
ridges as structural elements. We observed that the weakest point on the 
ridge is near the point of largest stress to curvature ratio. Thus
ridges which are used as support elements could be reinforced 
selectively at areas determined to be weak points through this analysis.

\section{Acknowledgements}
I'd like to thank Tom Witten for constant guidance and encouragement.
Sincere thanks also to Shankar Venkataramani for encouragement
and sharing of ideas, and to 
Eric Kramer and Alex Lobkovsky
for enlivening discussions. 

This work was supported in part by the National Science Foundation under 
Award number DMR-9975533 and by its MRSEC program under Award number 
DMR-9808595.

%


%

\end{document}